\documentclass[12pt,preprint]{aastex}

\usepackage{bm} 
\usepackage{amsmath}
\usepackage{amssymb}
\usepackage{bm}
\usepackage{epstopdf}
\usepackage{color}
\usepackage[normalem]{ulem}

\journal{Advances in Space Research}

\def\urms{u_{\rm rms}^\prime}

\def\avB{\langle {\hm B}\rangle }
\def\avBk{\langle { B_k}\rangle }

\def\av#1{\langle{#1}\rangle}

\def\turbB{{\hm B}^\prime}
\def\turbU{{\hm u}^\prime}
\def\turbJ{{\hm j}^\prime}

\def\bfalpha{\setbox2=\hbox{$\alpha$}
\hbox{{$\alpha$}\hskip-.97\wd2
{$\alpha$}\hskip-.97\wd2
{$\alpha$}\hskip-.97\wd2 {$\alpha$}}}

\newcommand{\dbdr}{\frac{\partial{\langle B_k\rangle}}{\partial{r}}}

\newcommand{\dbdtht}{\frac{\partial{\langle B_k\rangle}}{\partial{\theta}}}
\newcommand{\tildab}{\widetilde{b}}
\newcommand{\tildaa}{\widetilde{a}}

\begin{document}

\title{Characterisation of the turbulent electromotive force
and its magnetically-mediated quenching 
in a global EULAG-MHD simulation of solar convection}


\author{Corinne Simard, Paul Charbonneau and Caroline Dub\'e}
\affil{D\'epartement de Physique, Universit\'e de Montr\'eal, C.P. 6128 Succ.
Centre-ville, Montr\'eal, Qc H3C 3J7, Canada}
\email{corinne@astro.umontreal.ca}
\email{paulchar@astro.umontreal.ca}
\email{dube@astro.umontreal.ca}

\begin{abstract}
We perform a mean-field analysis of the EULAG-MHD millenium simulation of global magnetohydrodynamical
convection presented in \cite{passos2014}. The turbulent electromotive force operating
in the simulation
is assumed to be linearly related to the cyclic axisymmetric mean magnetic field
and its first spatial derivatives. At every grid point in the simulation's
meridional plane, this assumed relationship involves 27 independent tensorial
coefficients.
Expanding on \cite{racine2011},
we extract these coefficients from the simulation 
data through a least-squares minimization
procedure based on singular value decomposition.
The reconstructed $\hm\alpha$-tensor shows good agreement with that obtained
by \cite{racine2011}, who did not include derivatives
of the mean-field in their
fit, as well as with the $\hm\alpha$-tensor extracted by \cite{augustson2015}
from a distinct ASH MHD simulation. The isotropic part of the turbulent
magnetic diffusivity tensor $\hm \beta$
is positive definite and reaches values of
$5.0\times10^7\,$m$^2\,$s$^{-1}$ in the middle of the convecting fluid layers.
The spatial variations of
both $\alpha_{\phi\phi}$ and $\beta_{\phi\phi}$ component are well reproduced by
expressions obtained under the Second Order Correlation Approximation, with
a good matching of amplitude requiring a turbulent correlation time about
five times smaller than the estimated
turnover time of the small-scale turbulent flow.
By segmenting the simulation data into epochs of magnetic cycle
minima and maxima, we also measure $\alpha$- and $\beta$-quenching. 
We find the magnetic quenching of the $\alpha$-effect to be driven primarily
by a reduction
of the small-scale flow's kinetic helicity, with variations of the current
helicity playing a lesser role in most locations in the simulation domain.
Our measurements of turbulent
diffusivity quenching are restricted to the $\beta_{\phi\phi}$ component,
but indicate a weaker quenching, by
a factor of $\simeq 1.36$, than
of the $\alpha$-effect, which in our simulation drops by a factor of three 
between the minimum and maximum phases of the magnetic cycle.
 \\
\end{abstract}



\section{Introduction}

A proper understanding of the
physical mechanism(s) underlying solar dynamo action and regulating
the cycle's amplitude and duration are crucial components 
of long term prediction of space weather (also known as ``space climate''),
and of research on solar-terrestrial interaction in general
\citep{weiss2010}.
We are still a long way from physically-based prediction of solar
cycle characteristics, even though significant progress has been
made in recent years (for a recent review see \citealt{petrovay2010}).
Part of the difficulty lies with the fact that no concensus currently
exists as to the mode of operation of the solar cycle; the shearing
of the solar magnetic field by differential rotation is usually
considered as a key process, but what drives the regeneration of the
solar dipole moment remains ill-understood. Some dynamo models invoke
the electromotive force associated with turbulent convection, others
the surface decay of active regions (the Babcock-Leighton mechanism),
while others yet focus on
various rotationally-influenced
(magneto)hydrodynamical instabilities taking place immediately
beneath the base of the solar convection zone.
A survey of these different
types of dynamo models can be found in \citep{charbonneau2010}.
Such models make use
of geometrical and dynamical simplifications, most notably perhaps
the use of the so-called kinematic approximation, in which the
dynamical backreaction of the magnetic field on the inductive flows
is neglected or parametrized through largely ad hoc prescriptions.
Proper tuning of these ad hoc functionals and associated
model parameters can in many cases lead to cyclic behavior
showing reasonably solar-like variability patterns in the amplitude
and duration of magnetic cycles (see, e.g.,
\cite{karak2011, kitchatinov2012}
and references therein)

An alternate approach is made possible by global magnetohydrodynamical
simulations of solar convection, which recently have succeeded
in producing magnetic fields well-organized on large spatial
scales
and undergoing more or less regular polarity reversals
(\citealt{ brown2010,brown2011,ghizaru2010, kapyla2010, racine2011, kapyla2012, 
masada2013, beaudoin2013, passos2014,fan2014,augustson2015}).
There are
no active regions in such simulations (but do see \citealt{nelson2013,nelson2014}),
and therefore no Babcock-Leighton mechanism, but the turbulent electromotive
force associated with thermally-driven convection is captured in a dynamically
consistent manner at spatial and temporal scales resolved by the
computational grid. Evidence for the development of MHD instabilities
has also been found in some of these simulations (see \citealt{lawson2015,miesch2007}).

The availability of such simulation data allows to bridge the gap between
simplified kinematic models and MHD simulations of solar convection.
More specifically, the latter can be used to measure turbulent coefficients
usually specified in largely ad hoc fashion in the former. Of particular
interest is the turbulent electromotive force, its associated
$\alpha$-effect and turbulent diffusivity, and variations of these
as a function of the magnetic field strength. Such measurements can
assist in the interpretation of simulation results, and may help in
clarifying some puzzling differences in the characteristics of
cycles generated by simulations that are generally alike and differ
primarily in what one would have hoped are only computational and algorithmic
detail (see, e.g., \S 3.2 \citealt{charbonneau2014}). Moreover, mean-field
models incorporating source terms and physical coefficients derived
from numerical simulations can be useful in exploring long timescale
behaviors that remain unaccessible to full MHD simulations, due to limitations
in computing resources.

The aim of this paper is twofold.
First, we document and validate
a generalization of the least-squares minimization
technique introduced  by \cite{racine2011}(see also \citealt{brandenburg2002}) for extracting mean-field
coefficients from the output of global MHD simulations of solar convection.
Second, we use this methodology to measure the level of magnetically-mediated
quenching of the $\alpha$-effect and turbulent diffusivity operating in
the simulation.
Section \ref{sec:meanfield} presents a minimal overview of classical mean-field electrodynamics,
focusing on aspects necessary to properly frame the analyses to follow.
In section \ref{sec:SVD} we describe the least-squares minimization method used
to extract the $\hm \alpha$- and $\hm \beta$-tensors, and present the results
of this procedure in \S \ref{sec:alpha} and \ref{sec:beta}, applied to the ``millenium simulation'' described
in \cite{passos2014}. We also compare in \S \ref{sec:soca} the isotropic part of these
two tensors to reconstructions using analytical forms
obtained under
the second-order correlation approximation.
In sections \ref{sec:quenching} we turn
to an investigation of the magnetic suppression of the $\alpha$-effect
and turbulent diffusivity.
We close in \S \ref{sec:discussion} by summarizing our conclusions and discussing the limitation
of our analyses.

\section{Mean-field electrodynamics\label{sec:meanfield}}

The mathematical and physical underpinnings
of mean-field electrodynamics are well-covered
in many textbooks and review articles
(see, e.g., \citealt{moffatt1978,krause1980,ossendrijver2003,brandenburg2005,charbonneau2010}).
What follows is only a brief overview,
focusing on definitions and reformulations of the 
$\hm \alpha$- and $\hm \beta$-tensors on which the analyses presented
in this paper are based.
The starting point of classical mean-field electrodynamics
is the separation of
the magnetic field (${\bm B}$) and flow (${\bm u}$)
into a spatially large-scale, slowly varying mean
component, and a small-scale, rapidly varying
fluctuating component:

\begin{equation}
 \bm u = \langle \bm u \rangle + \bm u', \;\;\; \bm B = \langle \bm B \rangle + \bm
B^\prime~,
 \label{mean}
\end{equation}
where the prime quantities represent the fluctuating part and the brackets
$\langle ... \rangle $
denote an intermediate averaging scale over which the fluctuating parts vanish,
i.e., $\langle \bm u' \rangle=0$ and $\langle \bm B' \rangle=0$.
Inserting eq. (\ref{mean}) into the magnetohydrodynamical induction equation and
applying this averaging operator yields:

\begin{equation}
 \frac{\partial{\langle\bm B \rangle}}{\partial{t} } = \nabla \times \left(  \langle
\bm u \rangle \times \langle \bm{B} \rangle + \mathcal{\bm E} - \mathrm{\eta} \nabla \times \langle \bm{B}  \rangle \right)~,
 \label{induction}
\end{equation}
where
\begin{equation}
 \mathcal{ \bm E} = \langle \bm u^\prime \times \bm B^\prime \rangle~,
 \label{emf1}
\end{equation}
is the mean electromotive force ({\it emf}) due to the
fluctuation of the flow and the magnetic field, and
$\eta$ is the magnetic diffusivity. The next step is to develop
this {\it emf} in terms of the mean magnetic component and its derivatives.
Because we are working here with vector fields,
such a development is written as:

\begin{equation}
 \mathcal{\hm E} =  \hm a\avB + \hm b\nabla \avB + \textrm{higher order derivatives}~,
 \label{alphaserie} 
\end{equation}
where the tensors $\hm a$ and $\hm  b$ appearing in this expression
are assumed to depend only on the statistical properties
of the small-scale flow and field 
(see, e.g., \citealt{krause1980}). 
Truncation of the higher order derivatives is justified
provided a good separation of spatial and/or temporal
scale exists between the fluctuating velocity and magnetic fields
on one hand, and the large-scale magnetic and flow field
on the other.
The first term in the expansion involves a rank-two tensor capturing 
(among other effects) the
so-called $\alpha$-effect, which can act
as a source term in the mean-field equation (\ref{induction}). The second term in the series
is a rank-three tensor and embodies (among other effects)
the destructive action of turbulent
diffusion on the mean magnetic field
(see \citealt{radler1980, radler2000, radler2006}).
It is convenient ---and physically meaningful--- to separate out the symmetric
part of these tensors, so that with the higher-order terms neglected
eq.~(\ref{alphaserie}) can be rewritten as
\footnote{Note that the sign convention we use for $\hm\alpha$ differs from
\cite{schrinner2007}, who introduce minus signs on the $\hm\alpha$ and $\hm\gamma$ terms in both
eqs.~\ref{emf2} and \ref{eq:alphasym}; the final signs of the $\hm\alpha$ and $\hm\gamma$ components remains
the same under either convention.}:
\begin{equation}
  \mathcal{E} = \hm \alpha\cdot\avB +\hm\gamma\times\avB -\hm\beta\cdot(\nabla\times\avB) - \hm\delta\times(\nabla\times\avB) - \hm\kappa\cdot(\nabla\avB)^{sym} ~,
 \label{emf2}
\end{equation}
where 
\begin{equation}
 \alpha_{ij}={1\over 2}(a_{ij}+a_{ji})~,\qquad 
 \gamma_{k}=-{1\over 2}\epsilon_{kij}a_{ij}~.
 \label{eq:alphasym}
\end{equation}
The $\hm\alpha$-term now describes the classical $\alpha$-effect, which
in the present context
is related to the kinetic helicity of the unmagnetized flow, and the
vectorial quantity $\hm\gamma$ acts on the mean field as an
additional (pseudo)velocity known as turbulent pumping. 
The $\hm b$ tensor is separated into three components; a 
rank-two tensor $\hm\beta$, a vectorial quantity $\hm\delta$,
and a rank-three tensor
$\hm\kappa$. The first two can be interpreted as 
anisotropic contributions to the mean-field resistivity, and the last one
embodies other more complex influences of the mean field.
The $\hm\beta$-tensor is the symmetric part of the more general $\hm b$-tensor and
is defined as follows:
\begin{equation}
 \beta_{ij}={1\over 4}(\epsilon_{i\mu\nu} b_{j\mu\nu}+ \epsilon_{j\mu\nu} b_{i\mu\nu})~.
\end{equation}

Working in spherical polar coordinate ($r,\theta,\phi$) and assuming
that $\avB$ varies only weakly in space and by construction is
axisymmetric ($\partial{}/\partial{\phi}\equiv 0 $), the {\it emf} can be rewritten as:
\begin{equation}
 \mathcal{E}_m= \tildaa_{mk}\avBk + \tildab_{mkr}\dbdr  + \frac{\tildab_{mk\theta}}{r}\dbdtht~,
 \label{coefaxy}
\end{equation}
where $\tildaa$ and $\tildab$ are pseudo-tensors, which can be related
to the true tensors
$\hm a$ and $\hm b$ by introducing proper covariant differentiation
for the 
$\partial{\langle B_j\rangle}/\partial{x_k} $, so as to account for the
curvilinear nature of the (spherical) coordinate system and its
associated unit vectors.
Because our adopted averaging is a zonal average, all
$\partial{\langle B_j\rangle}/\partial{x_\phi} $ are zero, and consequently
only 18 out of the 27 components of the
$\hm \tildab$ pseudo-tensor are accessible
(see \cite{schrinner2007} for further details). In spherical geometry
we thus have:

\begin{subequations}
 \begin{equation}
  \alpha_{rr} = \tildaa_{rr} - \frac{\tildab_{r\theta\theta}}{r}~,  \label{eq:amodif1}\\
\end{equation}
\begin{equation}
  \alpha_{r\theta}=\alpha_{\theta r}= {1\over 2}(\tildaa_{r\theta } + \tildaa_{\theta r } +  \frac{\tildab_{rr\theta}}{r} -  \frac{\tildab_{\theta\theta\theta}}{r})~,   \\
\end{equation}
\begin{equation}
  \alpha_{r\phi} = \alpha_{\phi r}= {1\over 2}(\tildaa_{r \phi} + \tildaa_{\phi r} -  \frac{\tildab_{\phi rr}}{r})~, \\ 
\end{equation}
\begin{equation}
  \alpha_{\theta\theta} = \tildaa_{\theta\theta} + \frac{\tildab_{\theta r \theta}}{r}~,  \\
\end{equation}
\begin{equation}
  \alpha_{\theta\phi} = \alpha_{\phi\theta}= {1\over 2}(\tildaa_{\theta\phi } + \tildaa_{\phi\theta} +  \frac{\tildab_{\phi r\theta}}{r})~, \\
\end{equation}
\begin{equation}
 \alpha_{ \phi\phi} = \tildaa_{ \phi\phi}~.\label{eq:amodif6}
\end{equation}
\end{subequations}
Note here the appearance of $\tildab$-related contributions to the
$\hm\alpha$-tensor, a direct consequence of spatial derivatives acting
on unit vectors of the spherical coordinate system.
Similar expressions relating the components of
$\hm\beta$ to those of $\hm \tildab$ are given by
equations (12a) through (15k) in \cite{schrinner2007}, and are not
replicated in full here, except for the diagonal components:
\begin{equation}
 \beta_{rr}=-\frac{1}{2}\;\tildab_{r\phi\theta}~, \, \,\,\, \beta_{\theta\theta}=\frac{1}{2}\;\tildab_{\theta\phi r}~, \, \,\,\, \beta_{\phi\phi}=\frac{1}{2}(\tildab_{\phi r \theta} - \tildab_{\phi\theta r})~.
 \label{betadig}
\end{equation}
Note finally that the isotropic part of the $\hm\beta$-tensor, i.e.:
\begin{equation}
\beta\equiv {1\over 3}(\beta_{rr}+\beta_{\theta\theta}+\beta_{\phi\phi})~,
\label{betaturb}
\end{equation}
corresponds to the coefficient of turbulent diffusivity
introduced
in the vast majority of published mean-field 
and mean-field-like dynamo models
of the solar cycle, including those relying on inductive source terms 
distinct from the turbulent electromotive embodied in the $\hm\alpha$-tensor.

\section{Extracting the $a$-and $b$-tensors\label{sec:alphabeta}}

The numerical data used in what follows is taken from the EULAG-MHD
``millenium simulation'' described in \cite{passos2014}; (see also
\citealt{ghizaru2010,beaudoin2013,charbonneau2013,smolarkiewicz2013}).
This global simulation of thermally-driven MHD convection
spans 1600yr of simulated time, and generates an axisymmetric
large-scale magnetic field undergoing regular polarity reversals 
on a $\simeq 40\,$yr cadence.

A number of distinct approaches have been designed to extract the $\hm a$- 
and $\hm b$-tensors from the output of MHD turbulence simulations. 
The test-field method (\citealt{schrinner2007}, see also \citealt{kapyla2009})
solves a set of evolution equations for the turbulent fluctuations
${\hm B}^\prime$ produced by the kinematic action of the flow on a set
of imposed large-scale ``test magnetic field''. 
The turbulent {\it emf} is then calculated directly via eq.~(\ref{emf1}), and with the
test fields playing the role of the large-scale magnetic field $\avB$,
the tensorial components of the {\it emf} development (\ref{alphaserie})
can be obtained.

An alternate, direct approach is possible in the case of simulations which
generate autonomously a large-scale magnetic field, as in the EULAG-MHD
millenium simulation. Subtracting this
large-scale field from the total magnetic field yields ${\hm B}^\prime$,
and a similar procedure applied to the total flow provides ${\hm u}^\prime$,
at which point the turbulent {\it emf} is directly computed via eq.~(\ref{emf1}).
In the case of axisymmetric large-scale magnetic fields, the large-scale
component are defined through zonal averaging. This results, for
every spatial grid point $(r_b,\theta_c)$
in a meridional plane of the simulation, in a time series
of $\bm{\mathcal{E}}(r_b,\theta_c,t)$, which is linked to the corresponding
time series for the components of the mean field and its first order spatial
derivatives via eq.~(\ref{alphaserie}).
For example, in the case of the zonal component $\mathcal{E}_\phi$ this would read:

\begin{eqnarray}
 \label{eq:devxiphi}
 \mathcal{E}_\phi(r_b,\theta_c,t) &=& 
 \tildaa_{\phi r}\langle B_r\rangle(r_b,\theta_c,t) \nonumber \\ 
 & + &  \tildaa_{\phi\theta}\langle B_\theta\rangle(r_b,\theta_c,t) \nonumber \\ 
 & + & \tildaa_{\phi\phi}\langle B_\phi\rangle(r_b,\theta_c,t)  \\
 & + & \tildab_{\phi r r}\frac{{\partial \langle B_r\rangle}}{\partial r}
  +  \tildab_{\phi \theta r}\frac{{\partial \langle B_\theta\rangle}}{\partial r} 
  +  \tildab_{\phi \phi r}\frac{{\partial \langle B_\phi\rangle}}{\partial r} \nonumber \\
 & +&  {\tildab_{\phi r \theta}\over r}\frac{{\partial \langle B_r\rangle}}{\partial \theta} 
  +  {\tildab_{\phi \theta \theta}\over r}\frac{{\partial \langle B_\theta\rangle}}{\partial \theta} 
  +  {\tildab_{\phi \phi \theta}\over r}\frac{{\partial \langle B_\phi\rangle}}{\partial \theta} \nonumber ~.
\end{eqnarray}

Values for the $\tildaa(r_b,\theta_c)$ and $\tildab(r_b,\theta_c)$ components 
are then sought 
by least-squares minimization of the residual of the above expression.
At every grid point in the meridional plane, for each component of $\mathcal{\hm E}$
this involves 9 independent coefficients defining the linear combination
of the 9 time series of the mean magnetic field and its spatial derivatives 
on the RHS that best fit the {\it emf} time series on the LHS, for a grand
total of 27 unknown coefficients per spatial grid point.

\begin{figure}[!h]
\includegraphics[width=1.0\linewidth]{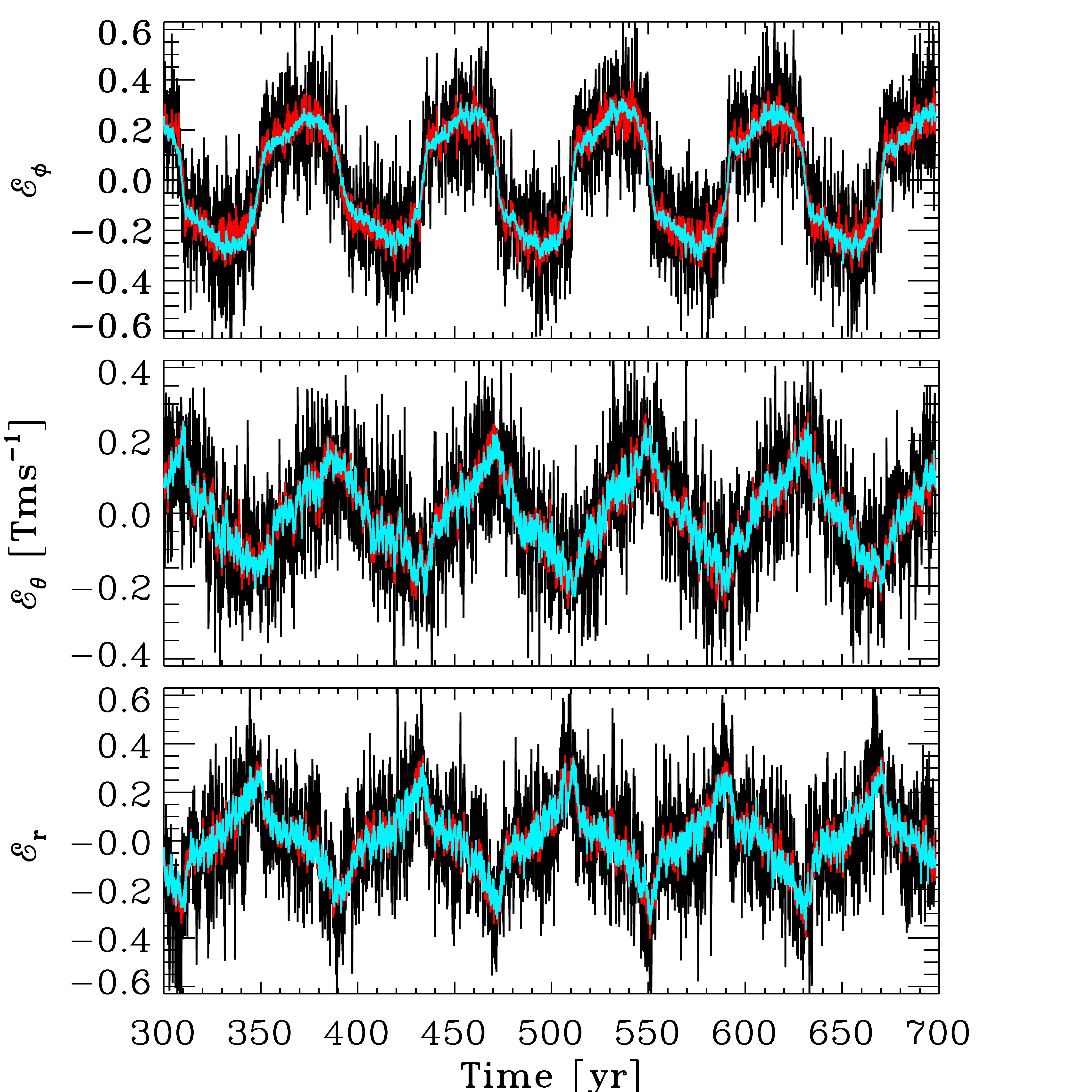}
\caption{Time series of the turbulent electromagnetic force (black)
 extracted directly from EULAG-MHD simulation output (viz.~eq.~(\ref{emf1}))
 in the course of 5 complete magnetic cycles.
The light blue time series show the {\it emf}
reconstructed using the original procedure 
of \cite{racine2011}, i.e., fitting only the 
$\hm\alpha$-tensor, while the red time series are the {\it emf} components
reconstructed from the expanded SVD procedure detailed below,
in which both 
$\hm\alpha$ and $\hm\beta$ are fitted to the {\it emf} (see text).
 }
\label{fig:EMF}
\end{figure}

Figure \ref{fig:EMF} shows a representative 400~yr segment of {\it emf} time series,
taken from the 1600~yr long EULAG-MHD millenium simulation described in
\citealt{passos2014},
used in the analyses to
follow. The components of the turbulent {\it emf} calculated directly from
the simulation output via eq.~(\ref{emf1}),
extracted at $45^\circ$ latitude in the N-hemisphere
at mid-convection zone depth, are plotted in black.
Even with the zonal averaging implied by eq.~(\ref{emf1}), the
{\it emf} components are quite noisy, but all shows a very well-defined
periodic signal. The colored time series are {\it emf} reconstructions
produced by the SVD-based least-square scheme described in what follows.

\subsection{The singular value decomposition procedure\label{sec:SVD}}

The procedure employed here to extract the $\hm a$ and $\hm b$-tensor
components
is a generalization of the method originally introduced by \cite{racine2011}.
The idea is
to take the {\it emf} and the mean (zonally-averaged)
magnetic field from the simulation output of EULAG-MHD and
seek time-independent numerical coefficients, specifically the
components of $\hm\tildaa$ and $\hm\tildab$, which provide the
best possible match to eq.~(\ref{coefaxy}). This defines an
optimization problem, which is tackled as a least-squares
minimisation using
Singular Value Decomposition (see \citealt{press1992}, Section~15.4).

For a given component $m=(r,\theta,\phi)$ of the {\it emf} at a
specific node of the computational grid $(r_b,\theta_c)$, we first define
the time-dependent 
functions $\mathrm{y}(t)$ and $X(t)$:
\begin{equation}
  \mathrm{y}(t) =\mathcal{E}_m(t,r_b,\theta_c)~,
\end{equation}
and 
\begin{equation}   
  X(t) =[ \langle B_k(t,r_b,\theta_c) \rangle, \,\partial_r\langle B_k(t,r_b,\theta_c) \rangle,\,\partial_\theta\langle B_k (t,r_b,\theta_c) \rangle ] \,\,\,\textrm{for}\,\,\, k=r,\theta,\phi~,
  \label{Xk1}
\end{equation}
where for each time step of the simulation output,
$X(t)$ at $(r_b,\theta_c)$ is a 9 components vector containing
the mean magnetic field and its derivatives, as appearing on the
RHS of eq.~(\ref{coefaxy}).
We then define the vector $\hm\varphi$ at fixed $m$ as:
\begin{equation}
  \varphi_j=[\tildaa_{mk}(r_b,\theta_c), \tildab_{mkr}(r_b,\theta_c),\tildab_{mk\theta }(r_b,\theta_c)]~,\,\,\,\textrm{for}\,\,\, k=r,\theta,\phi~,
\end{equation} 
containing the value of $\hm\tildaa$ for $j=1-3$ 
and $\hm\tildab$ for $j=4-9$ corresponding to a single component 
of the {\it emf}.
In this notation, the parameterization of the {\it emf} is written as:
\begin{equation}
  \mathrm{y}(t)=\sum\limits_{j=1}^9 \varphi_j X_j(t)~,
\end{equation} 
and the goal here is to find the 9 values of $\varphi_j$ that minimize
the least-squares merit function:
\begin{equation}
  \chi^2=\sum\limits_{i=1}^{N_t} \left[ \mathrm{y}(t_i)- \sum\limits_{j=1}^9 \varphi_j X_j(t_i) \right]^2  ~,
  \label{merit}
\end{equation}  
where $N_t$ is the number of time-steps $t_i$. 
The design matrix {\bf A} used in singular value decomposition procedure is
constructed as $ \mathrm{A}_{ij}=X_j(t_i)$. In the present context
this $N_t\times 9$ matrix depends on the large
scale magnetic field and its spatial derivatives in $r$ and $\theta$,
as per eq.~(\ref{coefaxy}). It can be decomposed as:
\begin{equation}
  \mathbf{A}= \mathbf{U \cdot w \cdot V^T} ~,
\end{equation} 
where $\mathbf{U}$ is an $N_t\times9$ orthogonal matrix, $\mathbf{w}$ is a $9\times9$ diagonal matrix
containing the so-called singular values,
and $\mathbf{ V}$ is a $9\times9$ orthogonal matrix. 
The solution
of $\hm \varphi$ is then given by;
\begin{equation}
  \hm \varphi= \mathbf{V \cdot w^{-1} \cdot U^T \cdot y} ~,
\end{equation} 
which is where the {\it emf} finally enters the problem, though the
$\mathrm{y}=(\mathrm{y}_1,\mathrm{y}_2,...\mathrm{y}_{N_t})$ time series.

This whole procedure is repeated three times
for $m=r,\theta$ and $\phi$ and at
all grid points in the meridional plane so as
to construct the spatial profiles of
all 9 components of the $\hm\tildaa$ pseudotensor, and of
the 18 components of the $\hm \tildab$ pseudotensor accessible
in a system axisymmetric on the large scales.

A great practical advantage of the SVD decomposition is that it also returns
the standard deviation ($\sigma$) via:
\begin{equation}
 \sigma^2_j = \sum\limits_{i=1}^{9}\left( \frac{\mathrm{V}_{ji}}{\mathrm{w}_{ii}}\right)^2~.
 \label{sigma}
\end{equation} 
For each component of $\hm\alpha$ and $\hm\beta$ computed, we can thus
recover an error estimate. In all results presented in what follows,
a measurement of a tensor component is deemed significant if it deviates
from zero by more than its associated standard deviation.
Note that $\sigma$ is a function of $\mathrm{ \mathbf{V}}$ and
$\mathrm{\mathbf{w}}$, both originating from the decomposition 
of the design matrix, which therefore implies that $\sigma$ is
set by fluctuations of the mean magnetic field and its derivatives, but
not by fluctuations of the  {\it emf} components.
Typically, the inclusion of the 
$\hm \beta$-tensor in the SVD procedure tends to increase the standard
deviation with respect to reconstruction fitting only the $\hm\alpha$-tensor,
as in \cite{racine2011}, because the time series of mean field spatial
derivatives tend to be noisier than those of the mean field components.

\subsection{Results for the $\boldmath{\alpha}$-tensor\label{sec:alpha}}

\cite{racine2011} extracted the $\hm\alpha$-tensor from the output
of an earlier EULAG-MHD simulation of much shorter duration, retaining
only the first term in the {\it emf} development given by eq.~(\ref{alphaserie}). Including now
the second term, proportional to first derivatives of the mean field
not only introduces additional explicit contributions to the $\hm\alpha$-tensor
components (see eqs.~(\ref{eq:amodif1})--(\ref{eq:amodif6})), but also alters 
the fit altogether. 
In either cases the reconstructed {\it emf} offers a very
good representation of the {\it emf} measured directly from the simulation
output via eq.~(\ref{emf1}) herein, as shown by the red and blue time series
superimposed on the measured {\it emf} on Figure \ref{fig:EMF}.
However, as we shall see presently, including the $\hm\beta$-tensor does reduce the rms
residual with respect to the measured {\it emf}.

\begin{figure}[!h]
\includegraphics[width=1.0\linewidth]{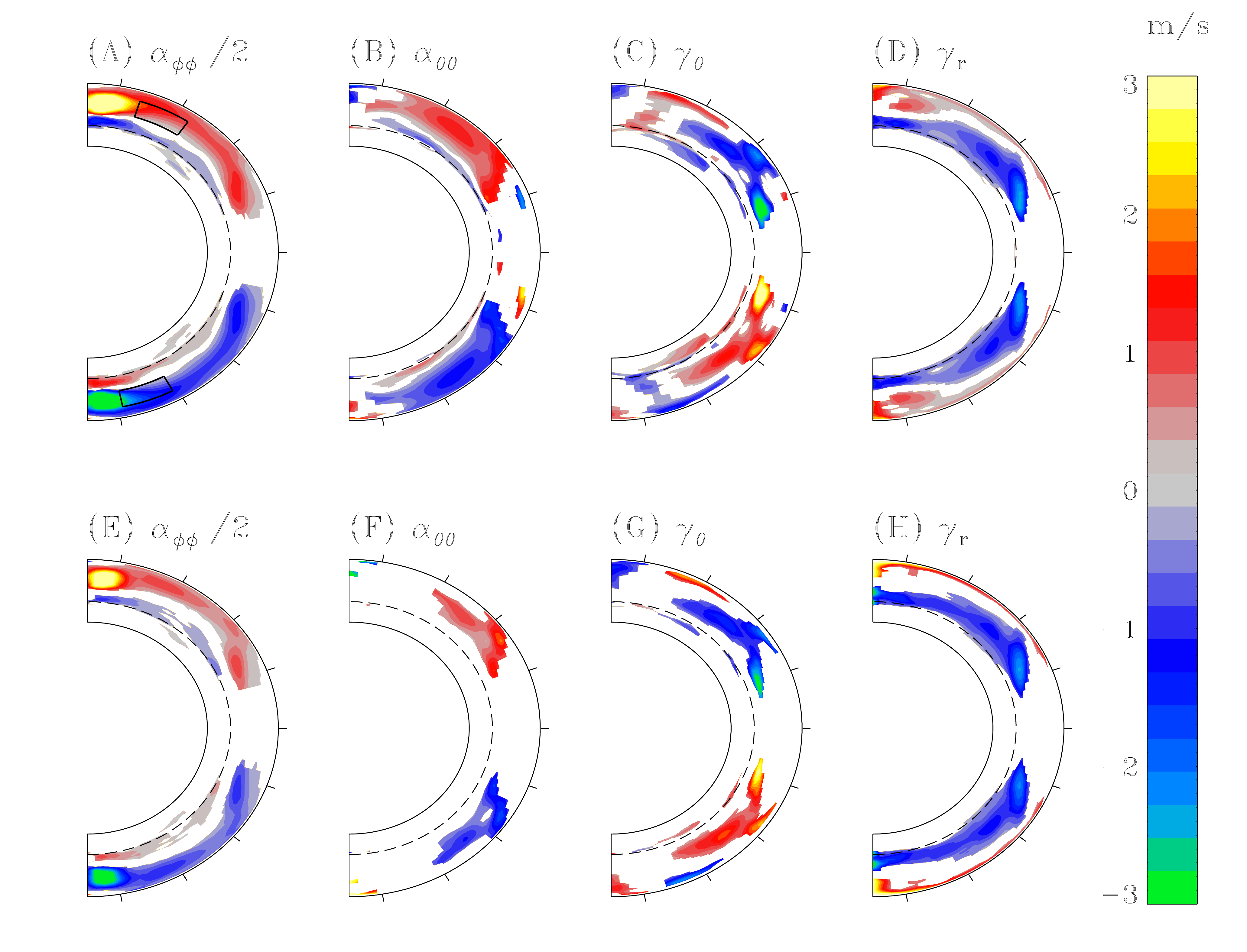}
\caption{Illustration of the $\alpha_{\phi\phi}$ (leftmost column)
and $\alpha_{\theta\theta}$ (second column) components,
together with the turbulent pumping components $\gamma_{\theta}$
(third column)  and $\gamma_{r}$ (rightmost column) as extracted from
the 1600$\,$yr long EULAG-MHD millenium simulation. The top row
shows results when only the first term is retained on the RHS of eq.~(\ref{alphaserie}),
while the bottom row shows the corresponding results when the first
two terms are retained.
A white mask is applied in regions of the meridional plane
where the signals is lower than
one standard deviation.
On each panels, the dash-line indicates the base of the convectively
unstable fluid layer, and the tickmarks on the outer boundary of each diagram
are drawn at intervals of $20^{\mathrm{o}}$.
The box drawn in panel (A) shows the integration domain used in the
$\alpha$-quenching analysis of \S \ref{sec:alphaquench} below.
 }
\label{fig:alpha}
\end{figure}

It will prove interesting to first quantify the differences between
the $\hm\alpha$-tensors, extracted with and without the $\hm\beta$-terms so as to
reassess the reliability and accuracy
of the \cite{racine2011} $\hm\alpha$-tensor extraction.
Such a comparison is presented on Figure \ref{fig:alpha}, for two of the
diagonal components of the $\hm\alpha$-tensor (two leftmost columns) and two
components of the turbulent pumping speed (two rightmost columns,
viz.~eq.~(\ref{eq:alphasym})). The top row shows results where only the $\hm a$ term is
retained in eq.~(\ref{alphaserie}), while in the bottom row both ${\hm a}$ and ${\hm b}$
are used in the SVD fitting procedure, as described in \S \ref{sec:SVD}.
A white mask is applied to show only regions of the meridional plane
where the measured tensor components deviate from zero by more than
one standard deviation.
Both sets of tensor component are morphologically quite similar,
the primary difference being a small reduction
in overall magnitude when
the ${\hm b}$ term is retained in the analysis, ranging from $\simeq 15$\%
for $\gamma_r$, up to $50$\% for $\alpha_{\phi\phi}$, this latter, larger difference
being dominated by variations in the polar regions. 

\begin{figure}[!h]
\includegraphics[width=1.0\linewidth]{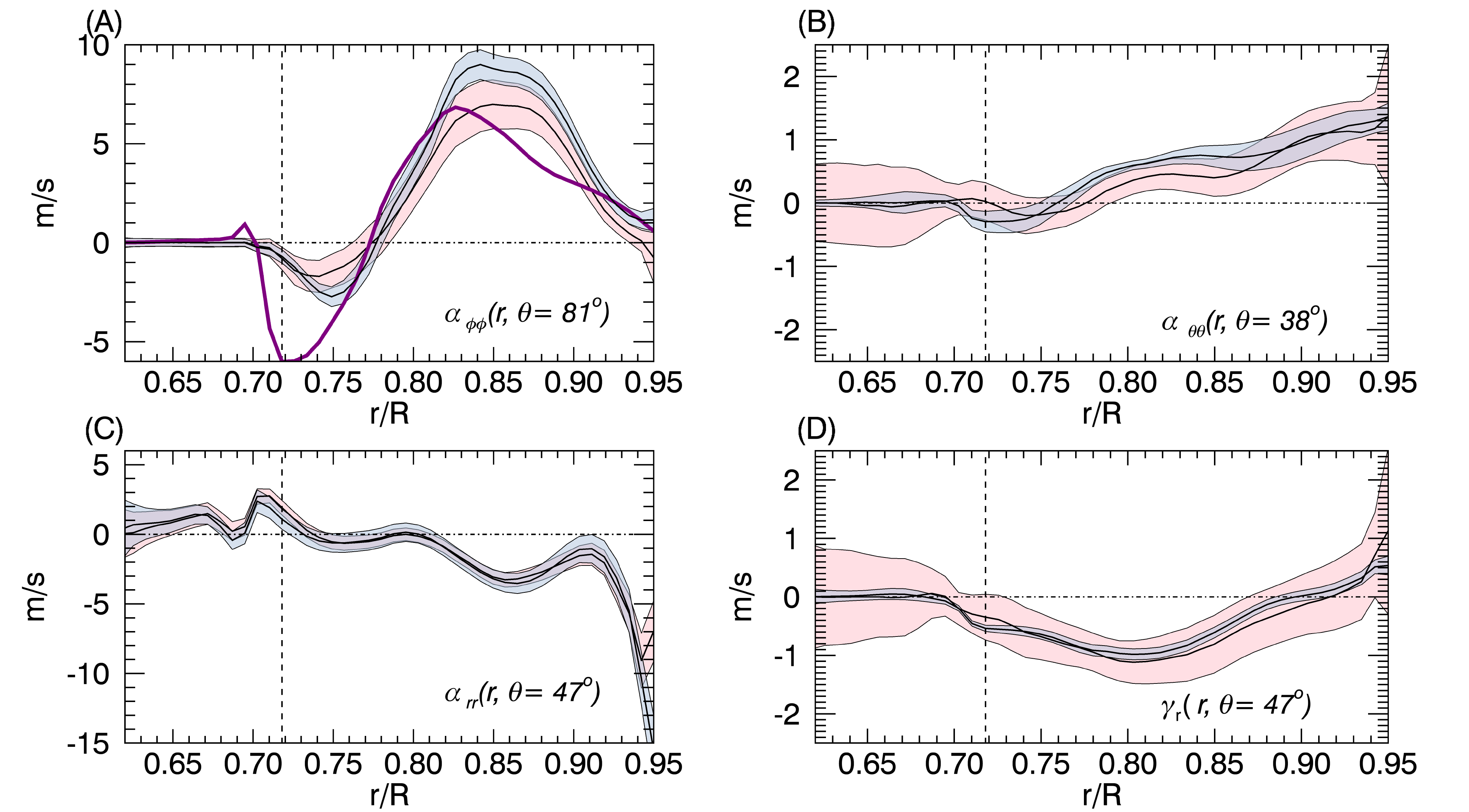}
\caption{ Radial cuts of four selected $\hm\alpha$-tensor components taken
at different latitudes 
 without (blue) and with (pink)
  the inclusion of $\hm\beta$ in the SVD fitting procedure. The
  colored area indicate the correponding 1$\,\sigma$ region of significance, and 
  the vertical dashed line the location of the core-envelope interface. 
  Panel (A) also includes (in blue) a radial profile of the SOCA $\alpha *$ (eq.~(\ref{eq:SOCAa}))
extracted at the same latitude and divided by a factor 4.5 (see text).
 }
\label{fig:alphacut}
\end{figure}

Figure \ref{fig:alphacut} shows radial cuts for four $\hm\alpha$-tensor components extracted at
different latitudes with and without the inclusion
of the $\hm\beta$-tensor. The sets of cuts usually stand
within each other's one-$\sigma$
standard deviation returned by the SVD fit,
the most significant difference being found with the 
$\alpha_{\phi\phi}$ component, which stand between 1 and $\simeq 2\sigma$
of each other at high latitudes (viz.~Fig.~\ref{fig:alphacut}A).

\subsection{Results for the $\boldmath{\beta}$-tensor\label{sec:beta}}

\begin{figure}[!h]
\includegraphics[width=1.0\linewidth]{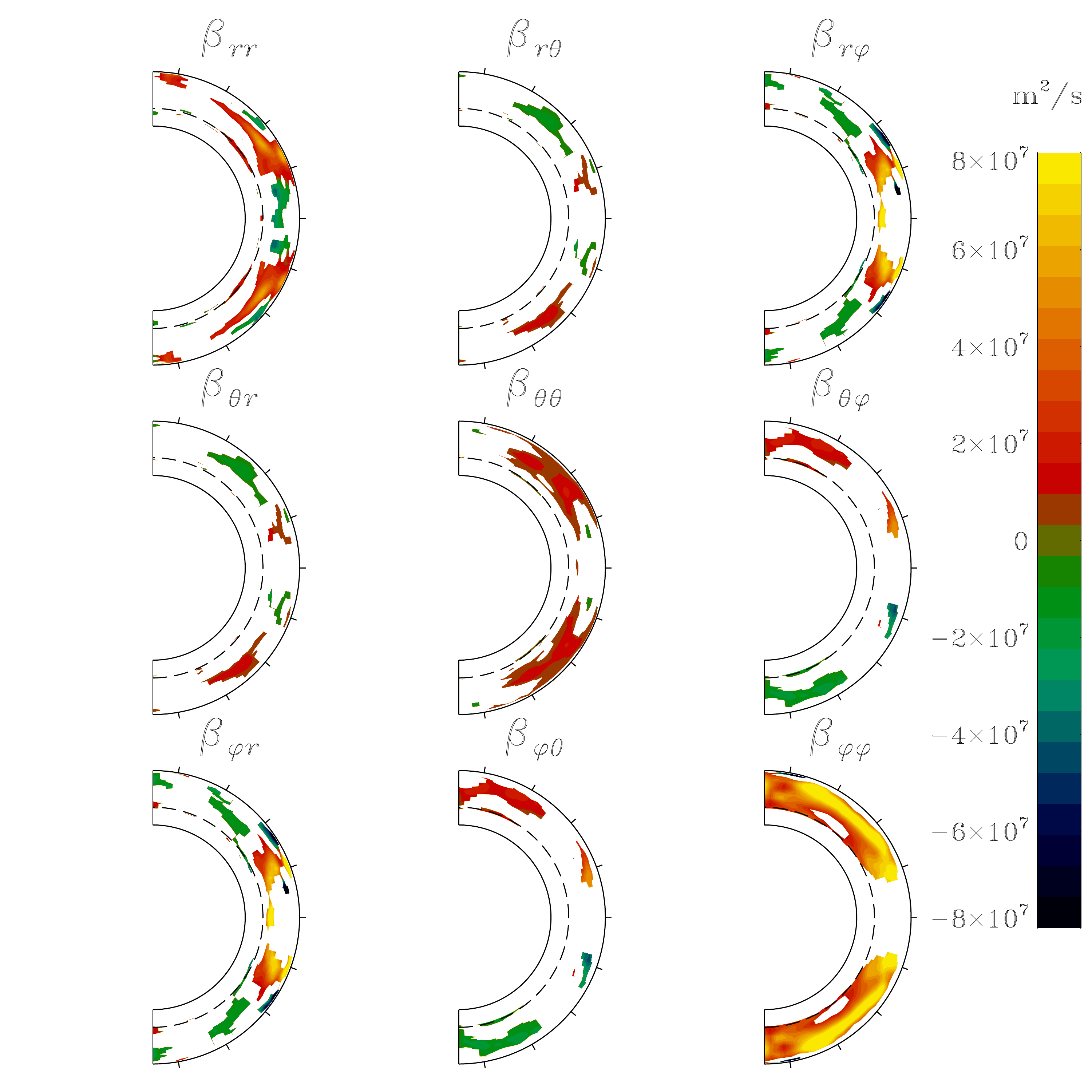}
\caption{Reconstruction of the 
$\hm\beta$-tensor
from the 18 accessible components of the pseudo-tensor $\hm\tildab$
extracted from the EULAG-MHD millenium simulation. The format is similar
to Fig.~\ref{fig:alpha}, with the dashed line indicating the base of 
the convectively
unstable fluid layers. A white mask applied to regions of the meridional
plane where the tensor components fall below the one-standard deviation level.
 }
\label{fig:beta9}
\end{figure}

Figure \ref{fig:beta9} shows the nine $\hm\beta$-tensor components reconstructed
via eqs.~(15d)---(15e) in \cite{schrinner2007}
from the $\hm\tildab$ components extracted using our SVD least-squares
minimization method. Once again a white mask is used to show only the
regions of the meridional plane where the tensor components exceed the
one standard deviation level, as returned by the SVD procedure.
As self-consistency check, we also applied a modified form of the SVD procedure, retaining only the $\hm\tildab$ coefficients, to the
residual $\bm{\mathcal{E}} - \hm\alpha \avB$ of a SVD extraction carried out only with the $\hm\tildaa$ terms, as in \cite{racine2011}.
The resulting $\hm\beta$-tensor closely resembles its counterpart returned by the complete extraction procedure described
in \S~\ref{sec:SVD} (viz. Fig.~\ref{fig:beta9}).

The tensor is noisier, and typically shows smaller significance regions 
than the $\hm\alpha$-tensor. Note however that the diagonal elements are
positive definite almost everywhere, which offers some confidence that
the results are physically meaningful. Moreover, the overall amplitudes
are in the range
$10^7$---$10^8\,\mathrm{m}^2/\mathrm{s}$, which is consistent with
other estimates of dissipation in similar EULAG simulations
(see, e.g., Strugarek et al., this volume).
The largest amplitude, peaking at
$\simeq 10^8\,\mathrm{m}^2/\mathrm{s}$,
are obtained for the $\beta_{\phi\phi}$ component,
with most of the domain returning a signal well above our one standard
deviation mask. This component thus dominates the diagonal, and will
be used preferentially in what follows when comparing to SOCA reconstructions
and measuring magnetically-mediated quenching.

\begin{figure}[!h]
\includegraphics[width=1.0\linewidth]{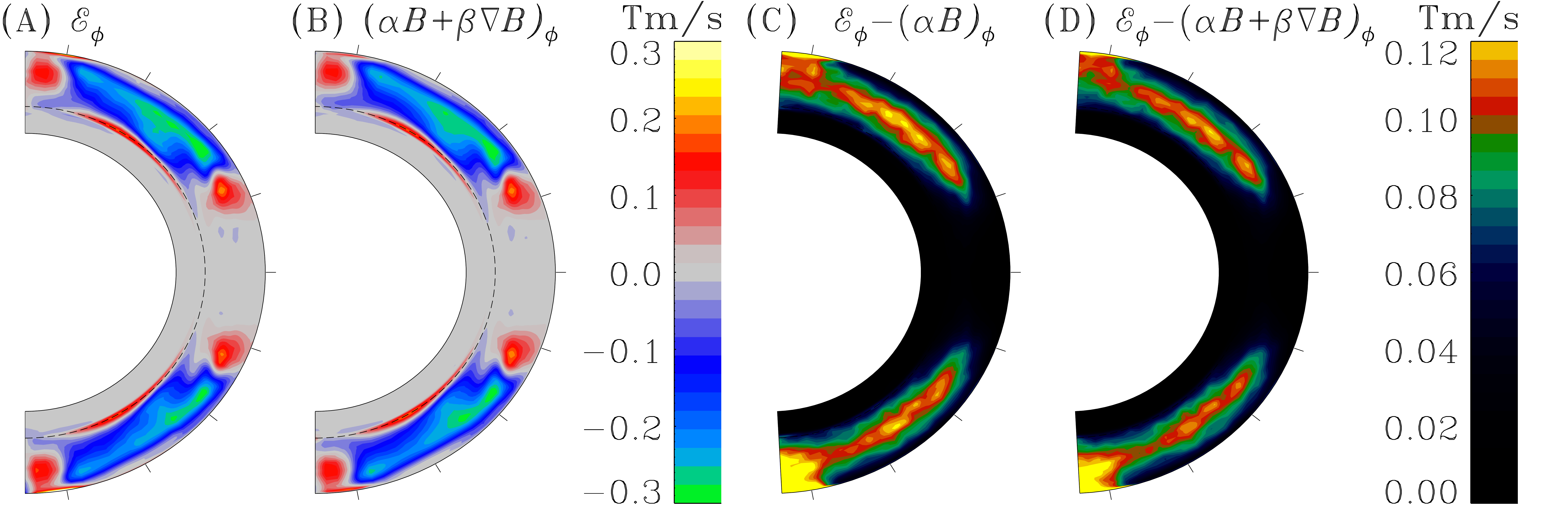}
\caption{ Mean azimutal turbulent electromotive force in a meridional plane
 for the concatenated set of cycle maximum 14 month-wide block as computed via eq.~(\ref{emf1}) in panel (A)
 and as reconstructed via eq.~(\ref{coefaxy}) in panel (B).
 Rms residuals averaged over the same concatenated set for the reconstructed {\it emf}
 where in the first (panel (C)) or both (panel (D)) term are retained in the SVD procedure.
 }
\label{fig:emfresidual}
\end{figure}

While many $\hm\beta$-tensor components lie below the 1$\sigma$ threshold in
extended portions of the meridional place, the rms residual
between the {\it emf} (Fig.~\ref{fig:emfresidual}A) and the reconstructed {\it emf}
(Fig.~\ref{fig:emfresidual}B) is reduced after the inclusion of the $\hm\beta$-tensor
in the SVD procedure.
This residual is plotted in Fig.~\ref{fig:emfresidual}D together
with the residual for a SVD fit using only the
$\hm\alpha$-tensor (Fig.~\ref{fig:emfresidual}C).
Comparison of those
two panels reveals a slight decrease of the residual when $\hm\beta$ is
included in the SVD procedure of the {\it emf}: averaged over the meridional
plane, the residual drops from $0.046 $ (Fig.~\ref{fig:emfresidual}C)
to $0.043\,\mathrm{Tm}/\mathrm{s}$ (Fig.~\ref{fig:emfresidual}D),
amounting to a 
6.5$\%$ decrease. This indicates that the slightly higher level of fluctuations
observed in the red time series on
Fig.~\ref{fig:EMF} does no result from a noisier reconstruction, but rather
from the fitting procedure capturing more of the physical variability present
in the extracted {\it emf} components.


\subsection{Comparison with SOCA\label{sec:soca}}

\begin{figure}[!h]
\includegraphics[width=1.0\linewidth]{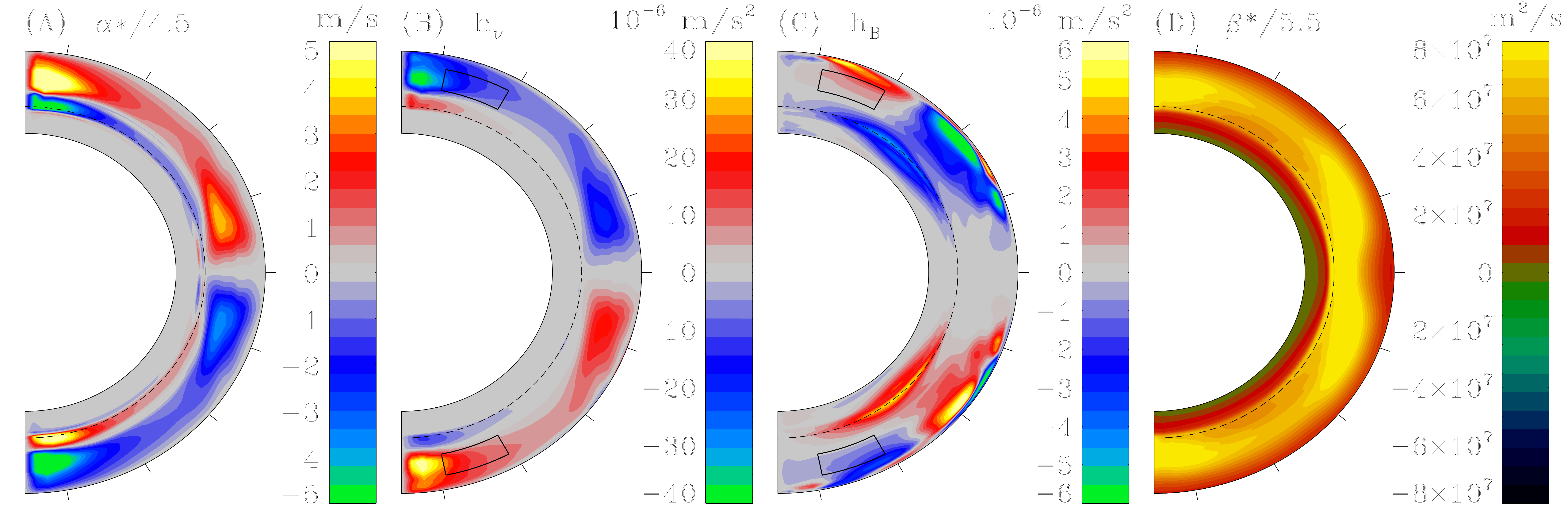}
\caption{ SOCA-based reconstructions of the isotropic part of the
$\hm\alpha$-tensor (in A) from the kinetic helicity (in B) as given
by eq.~(\ref{eq:SOCAa}); compare with Fig.~\ref{fig:alpha}A and E.
Panel C illustrate the current helicity as computed
by the second term on the RHS of equation (\ref{eq:nlalpha}),
note the different ranges of the color scales in (B) and (C).
Panel D shows the corresponding reconstruction
for the isotropic part of the $\hm\beta$-tensor via eq.~(\ref{eq:SOCAb});
compare with $\beta_{\phi\phi}$ at bottom right in Fig.~\ref{fig:beta9}.
The dashed line indicates the base of the convectively unstable fluid layers and
the boxes drawn in panels (B) and (C) shows again the integration domain used in the
following $\alpha$-quenching analysis of \S \ref{sec:alphaquench}.
}
\label{fig:SOCA}
\end{figure}

In the case of isotropic, homogeneous turbulence, the 
$\hm\alpha$ and $\hm\beta$-tensors simplify to $\alpha_{ij}=\alpha\delta_{ij}$
and $\beta_{ij}=\beta\epsilon_{ijk}$. The scalar coefficients $\alpha$ and
$\beta$ can be computed under the 
Second Order Correlation Approximation (SOCA) as:
%
\begin{equation}
 \alpha*=-\frac{\tau_c}{3} \langle \hm u'\cdot \nabla \times \hm u'  \rangle~, 
  \label{eq:SOCAa}
\end{equation}
and
\begin{equation}
 \beta*=\frac{\tau_c}{3}\langle (\bm u')^2 \rangle~,
 \label{eq:SOCAb}
\end{equation}
where $\tau_c$ is the coherence time of the turbulence,
and the terms in the averaging brackets on the RHSs
are, respectively, the mean kinetic helicity 
and the turbulent intensity of the small-scale flow component
(see, e.g., \citealt{ossendrijver2003,schrijver2009}).
Figure \ref{fig:SOCA}A and D show the corresponding $\alpha$ and $\beta$ coefficients,
assuming a coherence time equal to the turnover time of the
convective flow; the latter is estimated as $H_\rho/u^\prime$,
where $H_\rho$ is the density scale height and $u^\prime$ is
the rms small-scale flow speed.
The resemblance with the actual $\alpha_{\phi\phi}$, as extracted from the
simulation via the SVD least-squares method in Fig.~\ref{fig:alpha}E and Fig.~\ref{fig:alphacut}A, is
quite good:
both tensor components peak at polar latitudes,
show a secondary maximum at low latitude, and undergo
a sign change just above the core-envelope interface.
The main discrepancy resides in the
absolute magnitudes, with $\alpha*$ being larger than $\alpha_{\phi\phi}$
in its region of significance by an average factor of $\simeq 4.5$.

Figure \ref{fig:SOCA}D shows the corresponding
SOCA coefficient $\beta*$ computed via eq.~(\ref{eq:SOCAb}),
which is best compared to the
$\beta_{\phi\phi}$ component in the lower right of Fig.~\ref{fig:beta9}.
Here also the general spatial variations of the two quantities are similar,
the main difference being one of overall amplitude, with the SOCA expression
exceeding the extracted $\beta_{\phi\phi}$ by an average factor of $\simeq 5.5$
over its region of significance. That we recover a scaling factor almost
identical to that characterizing the ratio of $\alpha*$ to $\alpha_{\phi\phi}$
suggests that the discrepancy may lie with our choice of correlation
time $\tau_c$ in eqs.~(\ref{eq:SOCAa})---(\ref{eq:SOCAb}), indicating
in turn that equating $\tau_c$ to
the convective turnover time may be a large overestimate.
Interestingly, a short correlation time is one of the physical
regimes under which the SOCA approximation can be expected to hold
(\citealt{ossendrijver2003}).

\section{Magnetic quenching of the turbulent \textit{emf}\label{sec:quenching}}

In the nonlinearly saturated regime of dynamo action, one would expect
the Lorentz force associated with the magnetic field to impact
the inductive flows, including at the turbulent scales. Starting with the
pioneering study of \cite{pouquet1976}, this magnetic
quenching of the {\it emf} has by now been measured in a variety of MHD turbulence
simulations. \cite{karak2014} \S 1 give a good survey of these
various quenching measurements. In most cases what is being measured
is the suppression of the {\it emf} in response to the application of an
external large-scale magnetic field. 
This is also the case for the rotating convection simulations for which
\cite{karak2014} report quenching results: quenching is measured
with respect to the strength of imposed large-scale ``test-fields''.
One important exception is the
analyses of \cite{brandenburg2008}, who investigated quenching
of the $\alpha$-effect and turbulent diffusivity, both in full
tensorial form, in a cartesian box MHD simulation of helically forced
turbulence autonomously
generating a large-scale magnetic component. The EULAG-MHD
millenium
simulation introduced above also generates autonomously its own
large-scale magnetic field, and so offers the possibility to measure
directly the quenching of the {\it emf}, without the artificial introduction
of external large-scale field components.

\subsection{$\alpha$ Quenching\label{sec:alphaquench}}

\begin{figure}[!h]
\includegraphics[width=1.0\linewidth]{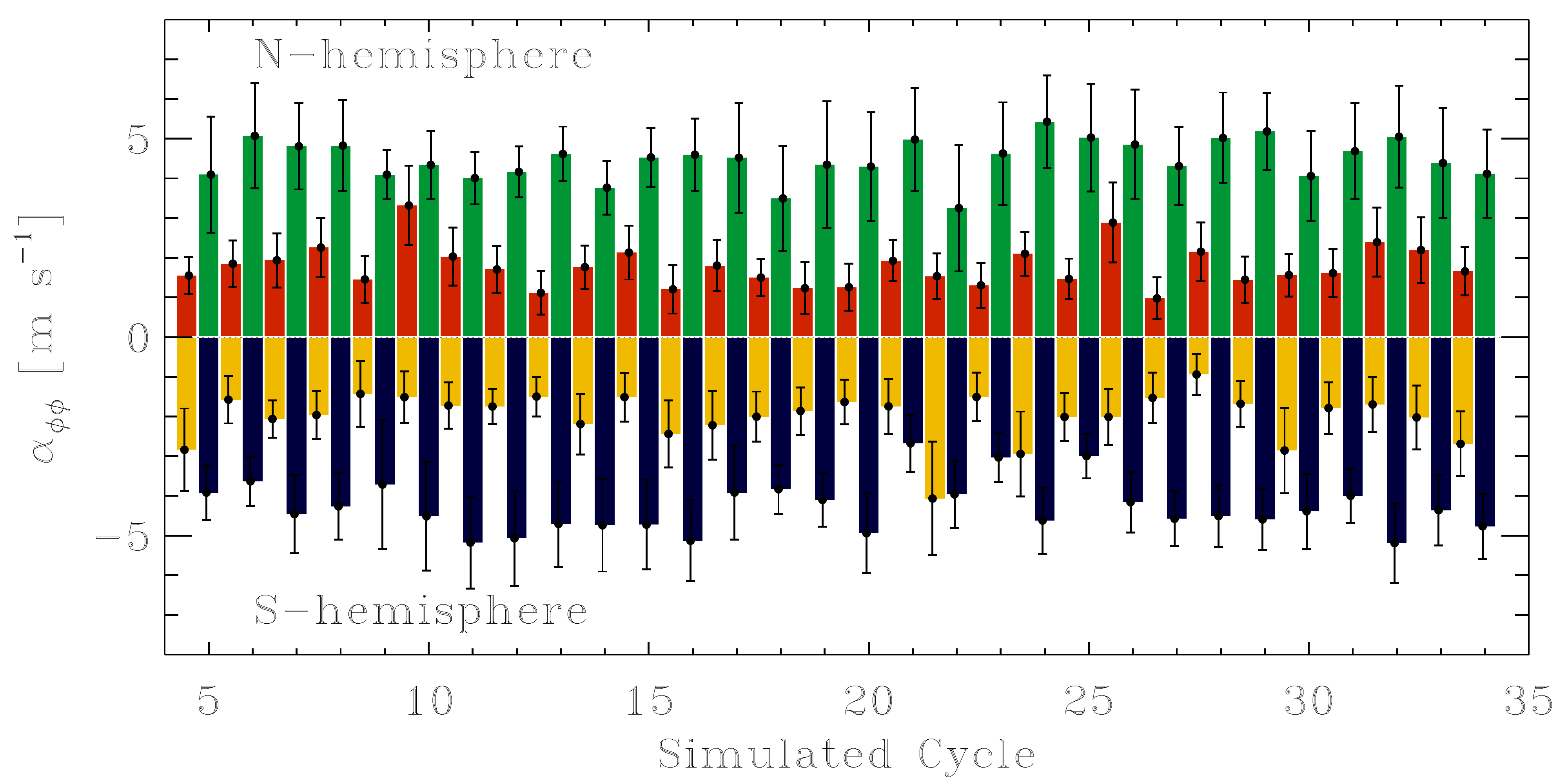}
\caption{Bar diagram showing the magnitude of the $\alpha_{\phi\phi}$
tensor component, averaged over the domain shown on Fig.~\ref{fig:alpha}A.
The top (bottom)
half of the diagram correspond to the Northern (Southern) hemisphere.
The SVD fitting procedure was applied here over 100-month wide segments
centered over successive cycle maxima (red and yellow) or minima (green 
and black). Error bars are estimated by averaging the standard deviation
over the same domain, assuming spatially uncorrelated statistics.
With only a few exceptions, cycle maxima show a level of $\alpha$-quenching
significantly exceeding the error bars.
}
\label{fig:alphabarNS}
\end{figure}

We first repeat the
SVD fitting procedure used to obtain the $\alpha_{\phi\phi}$ tensor component
plotted on Fig.~\ref{fig:alpha}A, this time over disjoint
100-month wide
temporal blocks (about one fifth of the half-cycle duration)
centered on epochs of cycle maxima and minima, the latter
determined on the basis of the time series of magnetic energy associated with
the large-scale magnetic component waxing and waning in the course
of the simulation. We opted to integrate the $\alpha_{\phi\phi}$ component
over the domain indicated on Figure \ref{fig:alpha}A. This selected
area is one where $\alpha_{\phi\phi}$ does not change sign,
has a magnitude much larger
than its standard deviation, and is located at high latitude, where the large-scale
dipole moment is building up (see Fig.~1B in \citealt{passos2014}). Henceforth,
unless explicitly stated otherwise, spatial averaging is always carried out
over this domain,
separately for each hemisphere. The results are shown on
Fig.~\ref{fig:alphabarNS}, in the form of bar charts, one bar per temporal
block, color-coded to indicate min/max cycle phase.
The error bars assigned to
each measurement
are obtained by integrating the standard deviation over the same
spatial domain as  $\alpha_{\phi\phi}$, assuming that fluctuations at
each grid point are uncorrelated. With only a few exceptions,
the $\alpha_{\phi\phi}$ component extracted from each of the 34 cycles
in the simulation
show a statistically significant difference between epochs of maxima
and minima. Qualitatively similar results are obtained for the other
components of the $\hm\alpha$-tensor.

\begin{figure}[!h]
\includegraphics[width=1.0\linewidth]{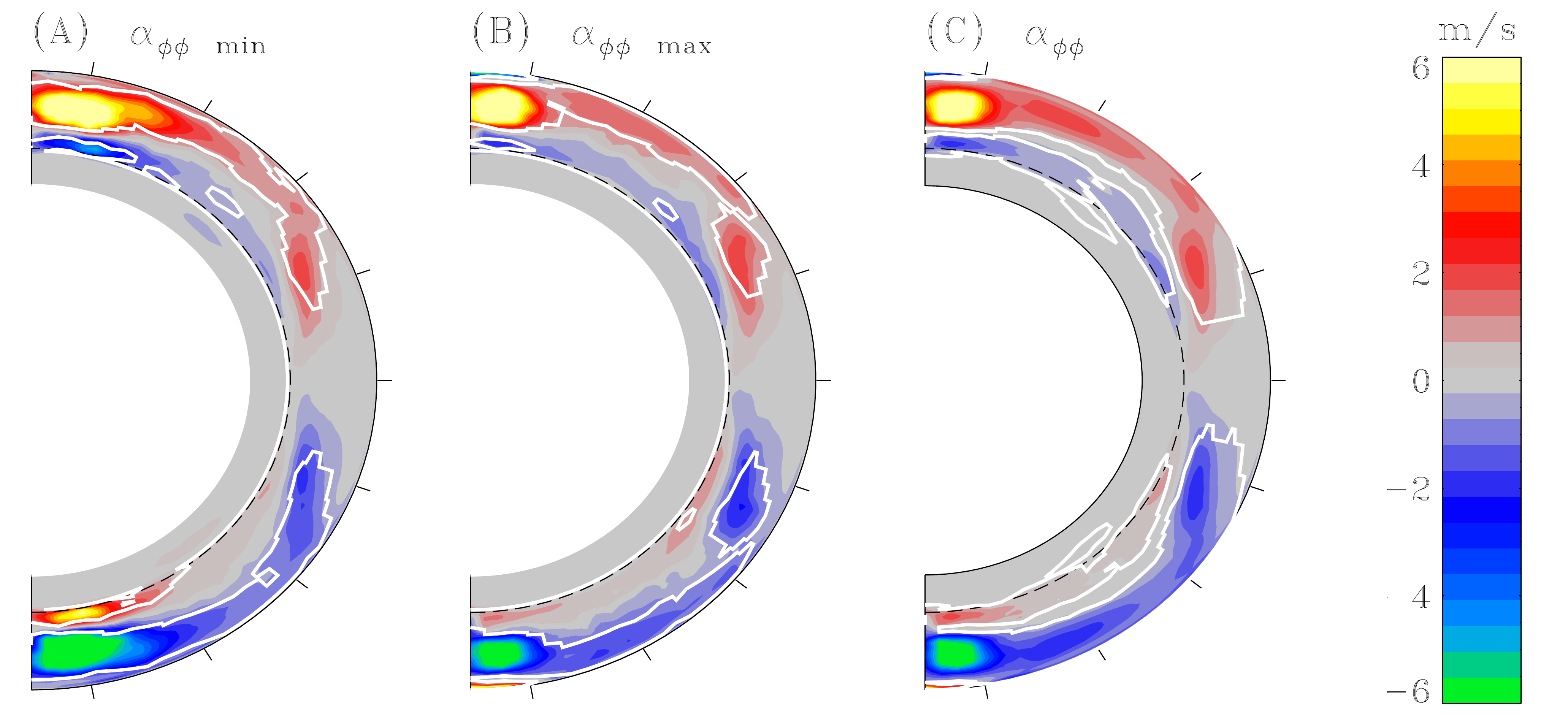}
\caption{The $\alpha_{\phi\phi}$ in the meridional plane, extracted
from (A) the concatenation of all ``minimum'' blocks on Fig.~\ref{fig:alphabarNS},
(B) the ``maximum'' blocks, and the full simulation in (C). White contours
delineate the regions of significance ($1\,\sigma$). 
}
\label{fig:alphamedcuts}
\end{figure}

Figure \ref{fig:alphamedcuts} shows, in  meridional planes, the spatial
profile of the $\alpha_{\phi\phi}$ tensor component for the concatenated set of 
cycle minimum 100 month-wide blocks in (A), cycle maximum blocks in (B), and
the full simulation in (C), for comparison purposes. In all cases 
white contours delineate the 1$\,\sigma$ significance
regions, as returned from the SVD algorithm. Comparing panels (A) and (B) reveals
that quenching of this tensor component involves an overall reduction of
its amplitude, leaving its spatial profile largely invariant.
The $\alpha_{\theta\theta}$ tensor component behaves similarly in this respect.

Next, we carry out a similar exercise, this time extracting the $\hm\alpha$-tensor
over successive 100-month long temporal block,
extending over the whole simulation with a 50\% overlap from block to block.
For each such block we average the $\alpha_{\phi\phi}$ component and
the magnetic energy over the same spatial domain as previously described.
The $\alpha_{\phi\phi}$ component shows
a clear decrease with magnetic energy, dropping from a mean value
$\simeq 4.4\,$m s$^{-1}$ at cycle minima, down to 
$1.8\,$m s$^{-1}$ at cycle maxima, amounting to a reduction by a 
substantial factor of three. Similar levels of quenching are observed with 
other $\hm\alpha$-tensor components, e.g., the averaged $\alpha_{r\theta}$ drops
from $1.5$ to $0.7\,$m s$^{-1}$ from cycle minimum to maximum.

\begin{figure}[!h]
\includegraphics[width=0.6\linewidth]{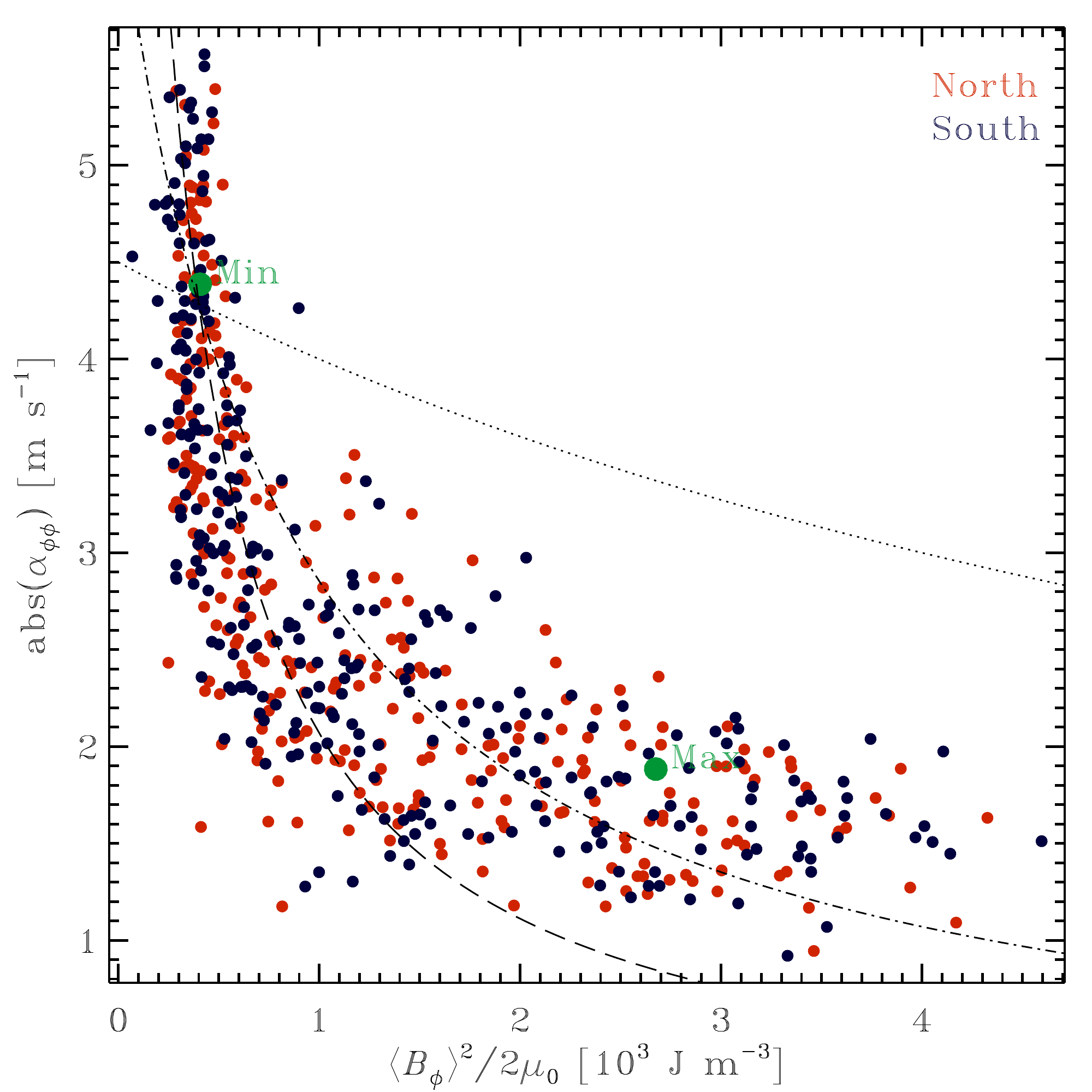}
\caption{Variation of the $\alpha_{\phi\phi}$ component versus magnetic
energy of the zonal magnetic component, both again averaged over the
domain shown on Fig.~\ref{fig:alpha}A. The SVD fit is carried out over
successive 100-month wide time blocks, with 50\% overlap between
successive blocks, and the magnetic energy is averaged similarly
in space and time. The larger green solid dots indicate the mean values for these
quantities at cycle maxima and minima, taken directly
from Fig.~\ref{fig:alphabarNS}.
$\alpha$-quenching is again quite obvious here, with
$\alpha_{\phi\phi}$ decreasing by a factor of $\sim 3$ over the range
of magnetic energy density sampled throughout the cycles.
The dotted curve shows the variation expected from the standard algebraic
quenching formula (eq.~(\ref{eq:aquench})), and
the dash-dotted and dashed curves result from the alternate strong
quenching expression
(viz.~eq.~(\ref{eq:squench})), with $R_m=10$ and $50$
respectively. 
}
\label{fig:alphaquenching}
\end{figure}

An $\alpha$-quenching
parametric formulae commonly used in mean-field
dynamo models is based on the assumption
that the $\alpha$-effect becomes suppressed once
turbulent fluid motions reach energy equipartition with the large-scale
magnetic field, i.e.,
\begin{eqnarray}
{1\over 2}\rho(\urms)^2={\avB^2\over 2\mu_0}~,
\end{eqnarray}
where $\mu_0$ is the permeability of free space and $\rho$ is the density of the plasma. 
The field strength at which this equality is satisfied defines the equipartition
field strength (hereafter denoted $B_{\rm eq}$).
Note that here the small-scale flow $\urms$ is extracted
from the simulation output of EULAG-MHD, and therefore is distinct from the
small-scale flow that would characterize a purely hydrodynamical simulation
operating in the same parameter regime. Such hydrodynamical flows
are typically used as baseline in many quenching analyses using externally-imposed
magnetic fields
(see e.g., \citealt{karak2014}).

The working
hypothesis embodied in eq.~(23) is most often introduced in mean-field models by adding
an explicit algebraic dependence on $\avB$ to the $\hm\alpha$-tensor components:
\begin{eqnarray}
\label{eq:aquench}
\alpha\to{\alpha_0\over 1+(\avB^2/B_{\rm eq}^2)}~,
\end{eqnarray}
where $\alpha_0$ is the magnitude of the $\hm\alpha$-tensor in the absence of
large-scale magnetic field.
This ad hoc expression obviously ``does the right thing'', in that it ensures
$\alpha\to 0$ as $\avB\gg B_{\rm eq}$. However, attempts to validate such expression against
MHD numerical simulations of forced helical flows
have instead lead to the alternate ``strong quenching'' expression
(\citealt{vainshtein1992}):
\begin{eqnarray}
\label{eq:squench}
\alpha\to{\alpha_0\over 1+R_m(\avB^2/B_{\rm eq}^2)}~,
\end{eqnarray}
where $R_m$ is the magnetic Reynolds number characterizing the flow.
With $R_m\sim 10^8$--$10^{10}$ in the
solar convection zone, $\alpha$-quenching then sets in at a magnitude
of $\avB$ four to five orders of magnitude below equipartition.
The difference between eqs.~(\ref{eq:aquench}) and
(\ref{eq:squench}) hinges on the fact that at high-$R_m$, the turbulent
flow first reaches
energy equipartition with $\turbB$, not $\avB$; eq.~(\ref{eq:squench}) then
follows from the scaling ratio
$\turbB/\avB\sim\sqrt{R_m}$, expected in the limit
$R_m\gg 1$ (see \citealt{cattaneo1996,hubbard2012}).
This catastrophic quenching is believed to reflect a cascade of magnetic
helicity to small scales, required under the constraint of total magnetic helicity
conservation (\citealt{brandenburg2001, field2002}). It
can be alleviated by allowing a flux
of helicity through the simulation boundaries 
(see, e.g., \cite{kapyla2008} and discussion therein).

In the simulation analyzed in this paper
$\rho=42\,$kg m$^{-3}$ and $\urms\simeq 20\,$m s$^{-1}$
in the middle of averaging domain used for the
$\alpha$-quenching analysis, which leads to a kinetic energy density
$e_k\simeq 8000\,$J m$^{-3}$, corresponding to an equipartition field strength
of $\simeq 0.15\,$T, in good agreement with the energy density of
the small-scale magnetic component averaged over the same subdomain.
Fig.~\ref{fig:alphaquenching} indicates that quenching is already well underway
at $\av{B_\phi}^2/2\mu_0\simeq 10^3\,$J m$^{-3}$. This suggests that
$\alpha$-quenching in our simulation is mediated primarily by the
small-scale magnetic field,
even though the magnetic Reynolds number
characterizing this simulation is only a few tens.

Attempts to fit the classical quenching expression given by
eq.~(\ref{eq:aquench}) to the simulation
data presented in Figure \ref{fig:alphaquenching} yield an extremely poor
fit (dotted line), due to the strong concavity of the observed trend.
Similar attempts using the strong quenching expression 
(\ref{eq:squench}) fare definitely better, although no single combination
of $\alpha_0$ and $R_m$ fits the simulation data well over its full
range.
The dash-dotted and dashed curves
on Fig.~\ref{fig:alphaquenching}
show the quenching predicted by eq.~(\ref{eq:squench}) for $R_m=10$
and $R_m=50$, respectively, with $\alpha_0$ adjusted to fit approximately
the mean $\alpha_{\phi\phi}$ value at cycle minima.
The rapid initial drop of $\alpha_{\phi\phi}$
is best reproduced by picking a higher $R_m$, but the flat, extended tail 
is better fit with a lower $R_m$. Note that
a magnetic Reynolds number of a few tens
is actually consistent with other estimates obtained for this simulation
by different means, e.g. from the turbulent spectrum,
and thus presents a form of internal consistency
with the strong quenching interpretation.

Whether in their weak or strong form, the algebraic quenching expressions
(\ref{eq:aquench}) and (\ref{eq:squench}) remain extreme simplifications
of the complexity of turbulent flow-field interactions.
The numerical simulations of \cite{pouquet1976} suggest that
for MHD turbulence with short coherence time, eq.~(\ref{eq:SOCAa})
should be replaced by:
\begin{equation}
\label{eq:nlalpha}
\alpha*=-{\tau_c\over 3}(\av{\turbU\cdot\nabla\times\turbU}
-\rho^{-1}\av{\turbJ\cdot\turbB})~.
\end{equation}
The first term in parentheses on the RHS of this expression is again the
kinetic helicity $h_v$, and the second is its magnetic equivalent,
namely the current helicity, where $\mu_0\turbJ=\nabla\times\turbB$,
a quantity closely related to the usual magnetic helicity.
Note that this magnetic contribution to the $\alpha$-effect
has a sign opposite to that of the kinetic contribution,
and reflects
the small-scale magnetic helicity 
will tend to counteract the kinetic helicity of the small-scale flow,
a general property
of flow-field interactions in the MHD limit.

We can take advantage of the fact that eq.~(\ref{eq:nlalpha})
offers a good representation of the $\alpha_{\phi\phi}$ component
extracted from the simulation (viz.~Fig.~\ref{fig:alpha}) to investigate
the physical origin of the $\alpha$-quenching measured in the simulation.
The formulation known as 
dynamical $\alpha$-quenching assumes that reduction of
the $\alpha$-effect takes place through
the growth of the magnetic term on the RHS of eq.~(\ref{eq:nlalpha}).
This growth is seen as an unavoidable consequence of magnetic helicity
conservation, which requires accumulation of magnetic helicity
of one sign at small scales,
if a large-scale magnetic component with helicity of opposite sign
is to be produced by turbulent
dynamo action (\citealt{brandenburg2001}).

\begin{figure}
\includegraphics[width=0.6\linewidth]{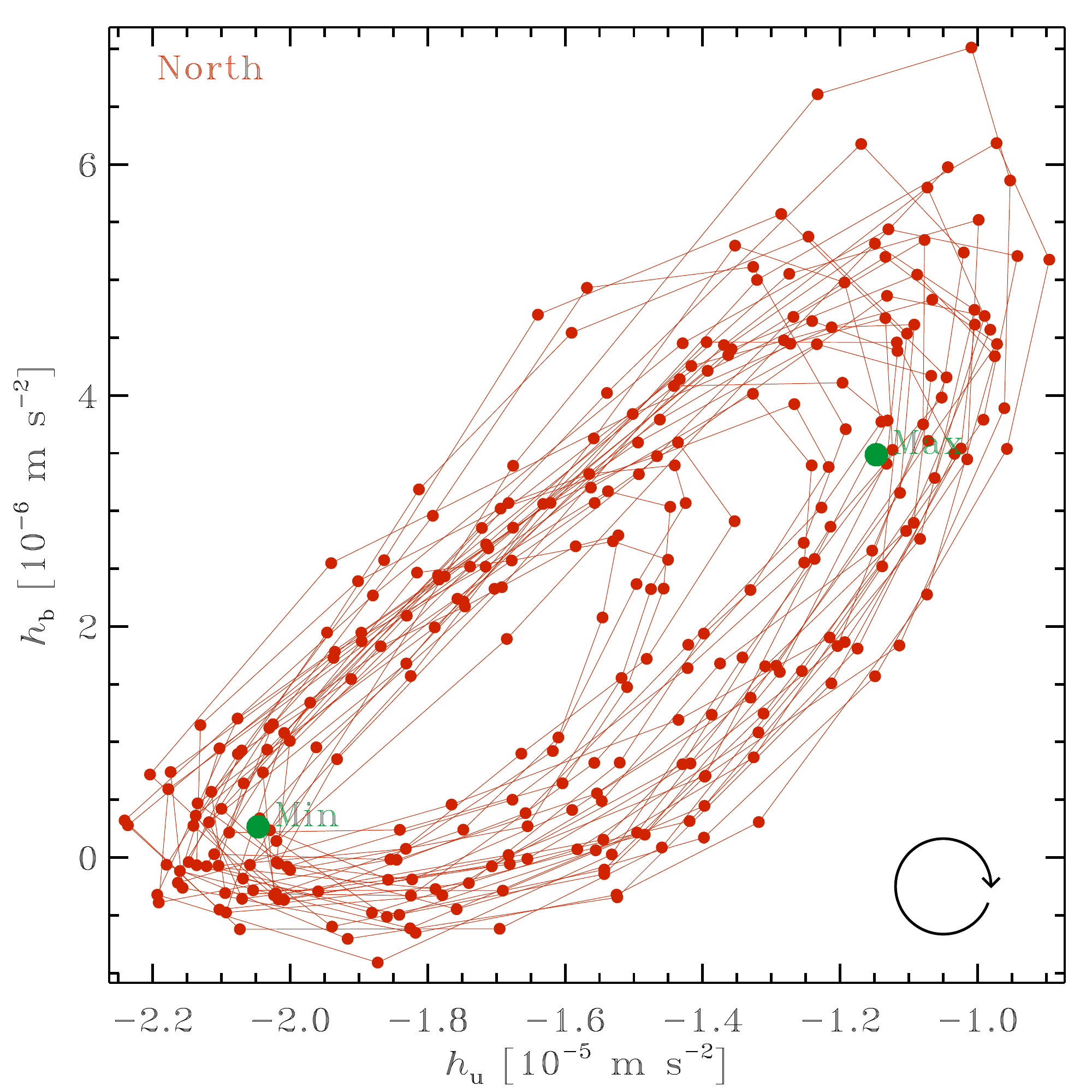}
\caption{Phase space portrait of the joint variations of the kinetic and
current helicities in the Northern hemisphere. As in previous Figures,
$h_v$ and $h_b$ are averaged over the domain
shown on Fig.~\ref{fig:SOCA}B and C and averaged over successive
100-month wide
temporal blocks with 50\% overlap (solid dots), with consecutive blocks
connected by a line segment. The trajectory runs clockwise on
this plot,
with mean values over cycle maxima and minima indicated by solid green dots.
}
\label{fig:alphaphasespace}
\end{figure}

Figure \ref{fig:alphaphasespace} shows the temporal
variations of the kinetic helicity $h_v$ and magnetic helicity $h_b$
over the course of the 34 cycles in the simulation, in the form
of a trajectory in the 2D phase space $[h_v,h_b]$. Both helicities
are averaged over the high latitude domain depicted on Fig.~\ref{fig:SOCA}B and C,
as well as in time, over 100-month wide temporal blocks overlapping by 50\%,
as on Figure \ref{fig:alphaquenching}.
The plot shows the trajectory associated with the 
Northern hemisphere, but the Southern hemisphere trajectory is similar,
except for being reflected about the origin. One full magnetic cycle corresponds
here to two clockwise circuits along the loop-like path, and the solid green dots
show the locii corresponding to the averages of maxima and minima
on Fig.~\ref{fig:alphabarNS}.

The cyclic growth of the current helicity $h_b$
from cycle minimum to subsequent maximum,
followed by a decrease to the next minimum, actually contributes an {\it increase}
of the $\alpha$-effect in the integration domain considered here
(box on Fig.~\ref{fig:SOCA}C).
This happens because in this region, the current helicity
has a sign opposite to the kinetic helicity, unlike in
most other regions of the domain, where the opposite situation
prevails and the current helicity opposes the kinetic helicity,
as per eq.~(\ref{eq:nlalpha}). This latter behavior
is generally consistent with the
picture of dynamical $\alpha$-quenching, according to which the
cascade of magnetic helicity to small-scale during the growth phase
of the cycle eventually leads to a saturation of the large-scale dynamo.
Nonetheless, here the increase of $\alpha$ mediated by the growing
current helicity is erased by a far more substantial variation of
the kinetic helicity, which drops
by almost a factor of two between minima and maxima. 
Repeating the same analysis with the integration region moved to lower
latitude yields different patterns, but in all cases the kinetic helicity
shows a reduction by factors in the range $\simeq 1.5$--2 between the minimum
and maximum phases of the magnetic cycle.

In view of 
the relative magnitudes of $h_v$ and $h_b$ (cf.~Fig.~\ref{fig:SOCA}B and C),
if eq.~(\ref{eq:nlalpha}) is taken at face value then one would conclude
that the cyclic variations of kinetic helicity dominates variations
of current helicity in quenching the $\alpha$-effect at most locations 
in our simulation domain.
This is a different pattern of quenching than
measured by \cite{brandenburg2008} in their
helically forced simulation, in which the kinetic helicity remained essentially
constant and the reduction of the $\alpha$-effect
could be traced to a corresponding increase of the 
current helicity. The difference is perhaps not surprising, since their
simulation is helically forced whereas in our case the forcing is thermal
and helicity is introduced through the action of the Coriolis force.

The EULAG-MHD ``millenium'' simulation providing the numerical data
used for all analyses presented
in this paper
achieves stability through implicit diffusivities associated with the
numerical advection scheme, which here is the same for the advection of
fluid velocity and magnetic field; in other words, here the 
magnetic Prandtl number is expected to be of order unity.
The rather complex variation of kinetic versus current helicity is
therefore unexpected, and must originate not with the dissipative
properties of the simulation, but rather with changes in the character
of the small-scale flows, presumably mediated by the large-scale
magnetic field and perhaps also time-varying large-scale flows.

\subsection{Diffusivity quenching}

Measurements of the quenching of turbulent diffusivity in MHD simulations
received comparatively less attention than $\alpha$-quenching. Section
1 of \cite{karak2014} provides again a good survey of these
measurements. Parametric diffusivity quenching, often akin to 
eq.~(\ref{eq:aquench}) herein, have been incorporated into some 
mean-field dynamo models (e.g., \citealt{rudiger1994}),
with goals as diverse as producing
interface dynamos (\citealt{tobias1996}), achieving
strong amplification to toroidal magnetic fields in the tachocline
(\citealt{gilman2005}), effecting the transition from diffusion-dominated
to advection-dominated regimes in flux transport dynamos (\citealt{guerrero2009}),
or producing a strong radial diffusivity gradient in the outer convective
envelope (\citealt{munoz-Jaramillo2011}).

\begin{figure}[!h]
 \includegraphics[width=1.0\linewidth]{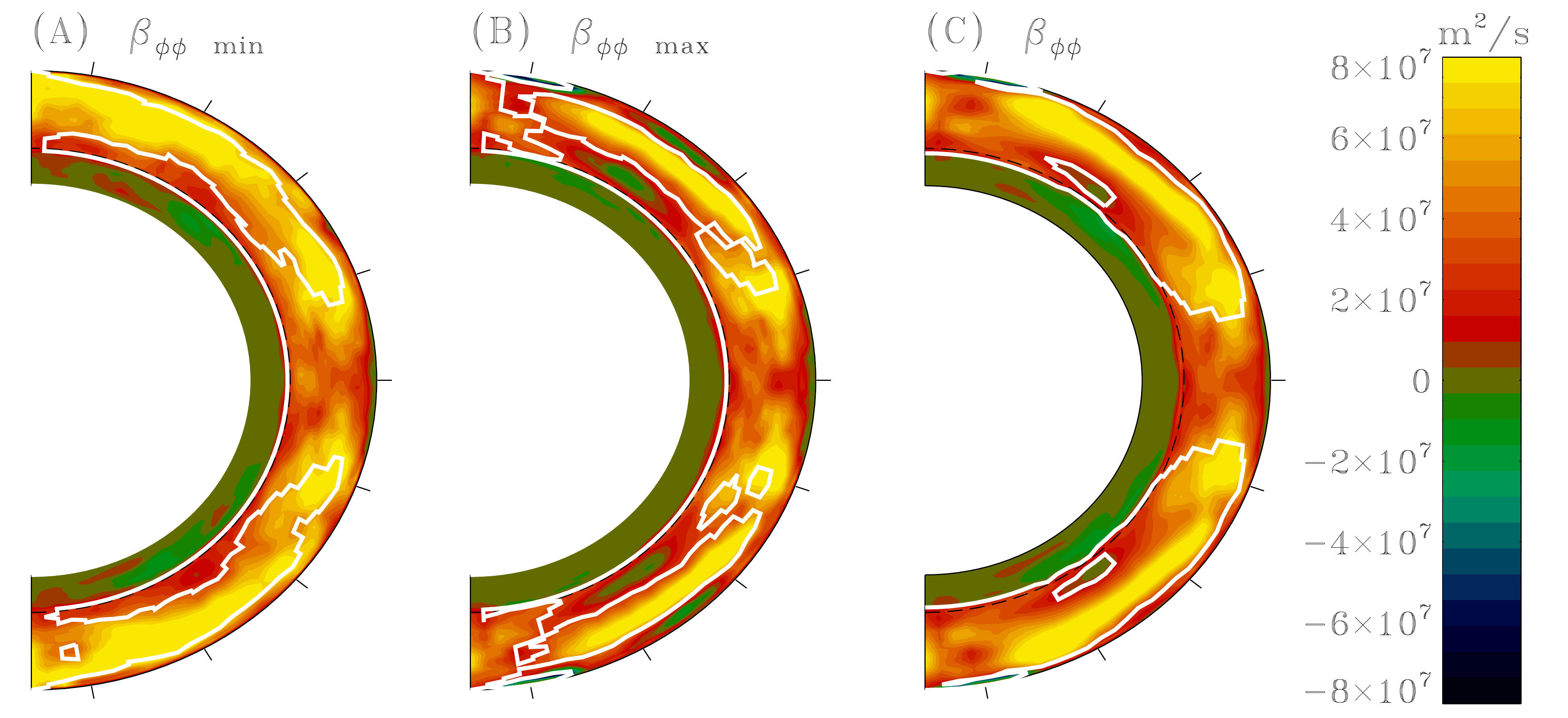}
 \caption{The $\beta_{\phi\phi}$ component extracted from the sets of
disjoint 100-month wide temporal segments centered either on (A) minima or (B)
maxima of the large-scale magnetic cycle. For comparison,
panel (C) shows the corresponding
plot for the whole simulation output, replicated from Fig.~\ref{fig:beta9}.
The white contour marks the level where
$\beta_{\phi\phi}$ is equal to the standard deviation
returned by the SVD
fitting procedure.}
 \label{fig:betaquench}
\end{figure}

In view of the $\hm\beta$-tensor measurements displayed on Fig.~\ref{fig:beta9},
we focus here only on the $\beta_{\phi\phi}$ component, for which a statistically significant determination
is obtained over $70\%$ of the meridional plane.
Following the same strategy just used for measuring $\alpha$-quenching,
we segment the simulation output into two sets of disjoint
100-month wide blocks centered respectively on magnetic energy minima and
maxima. We then repeat the SVD least-squares fit on the concatenation of
each of 
these two sets of segments. In this manner we generate
two ``average'' $\beta_{\phi\phi}$
components, respectively
characterizing epochs of maxima or minima of the magnetic cycle. The result
of this procedure is shown on panels A and B of Fig.~\ref{fig:betaquench},
with panel (C) replicating the $\beta_{\phi\phi}$ extracted from 
the full simulation run. In all three panels
the white contours marks the 1$\,\sigma$ region of significance as
returned by the SDV procedure. As expected, the corresponding level of noise for
$\beta_{\phi\phi}$ in epochs of minima and maxima are higher due to the reduced length of the simulation data.

Evidently, the overall amplitude of $\beta_{\phi\phi}$ is reduced
at times of cycle maxima, as compared to epochs of cycle minima. To
quantify this level of reduction we average $\beta_{\phi\phi}$ within
the statistically significant region located within the 1$\,\sigma$
contour, leading to values $6.0\times10^7\,$m$^2\,$s$^{-1}$ for minima, 
and $4.4\times10^7\,$m$^2\,$s$^{-1}$ for maxima, a $\simeq 36$\% 
reduction. In comparison, the
same average carried out for the $\beta_{\phi\phi}$ extracted
over the full simulation yields
$5.0\times10^7\,\mathrm{m}^2/\mathrm{s}$. The Min-to-Max ratio is
thus $\simeq 1.36$, much smaller than the reduction by a factor of
$\sim 3$ determined
for the components of the $\hm\alpha$-tensor (see Fig.~\ref{fig:alphaquenching}).
This is qualitatively consistent with other determinations of $\alpha$- and 
$\beta$-quenching in various types of 3D turbulent MHD simulations
(see \cite{karak2014}, \S 1, and references therein).

\section{Discussion and conclusion\label{sec:discussion}}

We presented in this paper a reconstruction of the
$\hm\alpha$ and $\hm\beta$-tensor characterizing the mean turbulent
electromotive force operating in a EULAG-MHD global simulation
of thermally-driven convection, specifically the ``millenium simulation''
presented in \cite{passos2014}. 
We generalized the original singular value decomposition-based
least-squares minimization
procedure of \cite{racine2011}, which was restricted to the $\hm\alpha$-
tensor, to include also the simultaneous reconstruction of 
the $\hm\beta$-tensor. 

Including the $\hm\beta$-tensor in the extraction procedure yields
results for the $\hm\alpha$-tensor that are quite similar to the earlier
extractions of \cite{racine2011}, who truncated the {\it emf} development
eq.~(\ref{alphaserie})
already at the first term. The primary difference is in the amplitude 
of the $\hm\alpha$-tensor components, which tend to be smaller
with the $\hm\beta$-tensor included. Interestingly,
\cite{augustson2015} report a similar level of variation when using their
implementation of the SVD least-squares extraction scheme on their
ASH simultion results, with the $\hm\beta$-tensor included or not in
the extraction process. Indeed, for the bulk of the convecting layers
the $\hm\alpha$-tensor extracted from \cite{augustson2015}'s ASH simulation
is quite similar to that characterizing our EULAG-MHD simulations
(cf.~Fig.~9 in \citealt{racine2011} to Fig.~13 in \citealt{augustson2015}),
keeping in mind that
our $\theta$ is latitude, which leads to a sign difference
for $\alpha_{(r\theta)}$, $\alpha_{(\theta\phi)}$, and $\gamma_\theta$).
This was not necessarily to be expected, considering that the two
simulations show some significant design differences, notably the presence
of stably stratified fluid layer underlying the convecting layer in
EULAG-MHD, distinct boundary conditions, very different
numerical and temporal resolutions and, more importantly, different numerical
diffusivities implicitly introduced at the smallest scales.

In a recent submission to ArXiv, \cite{warnecke2016} apply their preferred
test-field method, as well as a least-squares-based method analogous to that 
used in the present paper, to extract the $\hm\alpha$-tensors from
the output of a spherical wedge MHD simulations of solar convection.
Even though no error bars are provided for either method,
the authors claim that these
yield conflicting results (as per their Fig.~16), and on this basis conclude 
that the least-square-based method is ``unreliable'' (p 15) and ``incorrect'' 
(p 16). This conclusion stems from an additional a priori assumption 
being made, namely that the test-field results are by definition correct. 
Even if that were the case,
comparisons of this nature require quantitative error estimates
(cf.~their Fig.~16 to Fig.~\ref{fig:alphacut} herein), 
to assess whether the differences 
measured are statistically significant to begin with.

The $\hm\beta$-tensor we extract from our EULAG-MHD
millenium simulation is quite noisy,
with limited regions of 1$\,\sigma$ significance. Nonetheless,
the diagonal is clearly positive definite, with values 
$\sim 10^7$---$10^8\,$m$^2\,$s$^{-1}$,
commensurate with dissipation coefficient for similar EULAG simulations
estimated by other means (see Strugarek et al., this volume), as well
as with values typically used in mean-field dynamo models.

Our analysis of $\alpha$-quenching yields results in general
qualitative agreement
with the so-called strong quenching formulation, in which it is the
small-scale magnetic field that first reaches equipartition with turbulent
fluid motions and quenches the turbulent $\alpha$-effect.
Note however that the magnetic
Reynolds number characterizing our MHD simulation is relatively low,
of the order of a few tens, with the consequence that energy density
of the small-scale magnetic components exceeds that of the large-scale field
by less than a factor of ten in the region of the simulation domain
used for this quenching calculation.
Although cyclic variations of the small-scale current helicity are
measured in the simulation, 
the greater part of the $\alpha$-quenching
appears to result from a reduction of the turbulent kinetic helicity.
Approximately half of this decrease
is due to a drop in the rms turbulent flow speed, while the other half results
from a decrease in the alignment of the small-scale flow with respect 
to it vorticity vector.
These results indicate that, at least in this simulation,
quenching of the $\alpha$-effect is a fully magnetohydrodynamical phenomenon,
finding
its roots in magnetically-mediated changes in the patterns of turbulent 
convection (on this latter point see also \cite{cossette2016},
submitted to ApJ).

Independently of the applicability (or lack thereof) of the SOCA expressions for
the $\hm\alpha$-tensor,
our analysis indicates that even in the minimum phase of the magnetic cycles our
$\alpha$-effect shows a strong dependence on magnetic energy, indicating
that significant magnetic quenching is acting already then. 
Our $\alpha$-effect
evidently operates in a strongly nonlinear regime at all phases of the large-scale
magnetic cycles unfolding in the simulations, as seems to also be the case in 
the simulations of \cite{kapyla2012,kapyla2013}.

We also carried out measurements of diffusivity quenching, limited
at this point in time to the $\beta_{\phi\phi}$ component. We find
the turbulent diffusivity to suffer much less quenching than the 
$\alpha$-effect, specifically $\simeq 36\%$ versus a factor of $\simeq 3$
for $\bfalpha$.
This much weaker magnetic quenching of the turbulent diffusivity
is in qualitative agreement with the findings of
\cite{brandenburg2008}, even though the simulation analyzed by
these authors used a geometrical setup and turbulent forcing
quite different from ours.

Under a strictly scriptural interpretation of mean-field electrodynamics, 
the $\hm\alpha$- and $\hm\beta$-tensors are fundamentally linear,
hydrodynamical quantities which characterize
the inductive properties of a flow unaffected by the presence of magnetic
fields at any scale.
Here, in contrast, we are fitting eq.~(\ref{coefaxy}) to numerical data taken 
from a MHD simulation having reached its nonlinearly saturated stage.
The $\hm\alpha$- and $\hm\beta$-tensor extracted in this manner are thus
not, a priori, the same physical objects. They 
must be interpreted as coefficients quantifying
an empirical parametrization of the nonlinearly saturated
turbulent electromotive force in terms
of the large-scale magnetic field and its derivatives. 
That they do so in a meaningful manner is supported by the fact
that upon being inserted in a conventional axisymmetric kinematic
$\alpha^2\Omega$ mean-field model, they produce a large-scale
magnetic field exhibiting a spatiotemporal evolution resembling
reasonably well that observed in the original MHD simulation
from which the tensors are extracted (see \S 3.2 in \citealt{simard2013}).

More surprising is the fact that the isotropic parts of the $\hm\alpha$
and $\hm\beta$-tensors extracted from the simulation
show fairly good agreement with their linear forms,
as computed under the second order correlation approximation. The most
prominent discrepancy, namely the overall amplitude of the (isotropic)
SOCA $\alpha *$ and $\beta *$ being larger by a factor of about 5, may hold a clue
as to why the general spatial form of the tensors are otherwise so similar.
The SOCA estimates require the specification of the correlation time of
the turbulence flow ($\tau_c$ in eqs.~(\ref{eq:SOCAa}),
(\ref{eq:SOCAb}) and (\ref{eq:nlalpha})), a quantity notoriously difficult to extract from
numerical simulations. As a first cut we simply followed \cite{brown2010}
and set $\tau_c$ equal to the mean turnover time $\tau_u$ of the small-scale
flow, the latter estimated from the density scale height $H_\rho$ and rms
turbulent flow speed $u^\prime_{\rm rms}$;
in other words, we have assumed the Strouhal
number ${\rm St}=\tau_c/\tau_u$ to be unity.
Matching the amplitude of the SOCA coefficient
to those of the tensors extracted from the simulation requires
reducing the Strouhal number by a factor of $\simeq 5$, which then puts
us in the regime ${\rm St}<1$, in which the SOCA approximation
is actually expected to hold (\citealt{pouquet1976}).
We view this as an encouraging indication of
internal consistency in our analysis and interpretation, notwithstanding
the fact that the results presented here
pertain to a single simulation carried out at relatively low Reynolds
number, and that our turbulent diffusivity results remain at this writing
limited in scope.

We wish to thank Antoine Strugarek for useful discussions, and two
anonymous referees for some useful comments and suggestions.
This work was supported by the Natural Sciences and Engineering
Research Council of Canada, the Fond Qu\'eb\'ecois pour la Recherche -- Nature
et Technologie, the Canadian Foundation
for Innovation, and time allocation on the computing infrastructures of 
Calcul Qu\'ebec, a member of the Compute Canada consortium. 
CS is also supported in part through a graduate fellowship from FRQNT/Qu\'ebec.



%
%



\bibliography{references}

\begin{thebibliography}{53}
\expandafter\ifx\csname natexlab\endcsname\relax\def\natexlab#1{#1}\fi

\bibitem[{{Augustson} {et~al.}(2015){Augustson}, {Brun}, {Miesch}, \&
  {Toomre}}]{augustson2015}
{Augustson}, K., {Brun}, A.~S., {Miesch}, M., \& {Toomre}, J. 2015, \apj, 809,
  149

\bibitem[{{Beaudoin} {et~al.}(2013){Beaudoin}, {Charbonneau}, {Racine}, \&
  {Smolarkiewicz}}]{beaudoin2013}
{Beaudoin}, P., {Charbonneau}, P., {Racine}, E., \& {Smolarkiewicz}, P.~K.
  2013, \solphys, 282, 335

\bibitem[{{Brandenburg}(2001)}]{brandenburg2001}
{Brandenburg}, A. 2001, \apj, 550, 824

\bibitem[{{Brandenburg} {et~al.}(2008){Brandenburg}, {R{\"a}dler},
  {Rheinhardt}, \& {Subramanian}}]{brandenburg2008}
{Brandenburg}, A., {R{\"a}dler}, K.-H., {Rheinhardt}, M., \& {Subramanian}, K.
  2008, \apjl, 687, L49

\bibitem[{{Brandenburg} \& {Sokoloff}(2002)}]{brandenburg2002}
{Brandenburg}, A., \& {Sokoloff}, D. 2002, \gafd, 96, 319

\bibitem[{{Brandenburg} \& {Subramanian}(2005)}]{brandenburg2005}
{Brandenburg}, A., \& {Subramanian}, K. 2005, \physrep, 417, 1

\bibitem[{{Brown} {et~al.}(2010){Brown}, {Browning}, {Brun}, {Miesch}, \&
  {Toomre}}]{brown2010}
{Brown}, B.~P., {Browning}, M.~K., {Brun}, A.~S., {Miesch}, M.~S., \& {Toomre},
  J. 2010, \apj, 711, 424

\bibitem[{{Brown} {et~al.}(2011){Brown}, {Miesch}, {Browning}, {Brun}, \&
  {Toomre}}]{brown2011}
{Brown}, B.~P., {Miesch}, M.~S., {Browning}, M.~K., {Brun}, A.~S., \& {Toomre},
  J. 2011, \apj, 731, 69

\bibitem[{{Cattaneo} \& {Hughes}(1996)}]{cattaneo1996}
{Cattaneo}, F., \& {Hughes}, D.~W. 1996, \pre, 54, 4532

\bibitem[{{Charbonneau}(2010)}]{charbonneau2010}
{Charbonneau}, P. 2010, \lrsp, 7

\bibitem[{{Charbonneau}(2014)}]{charbonneau2014}
---. 2014, \araa, 52, 251

\bibitem[{{Charbonneau} \& {Smolarkiewicz}(2013)}]{charbonneau2013}
{Charbonneau}, P., \& {Smolarkiewicz}, P.~K. 2013, \science, 340, 42

\bibitem[{{Cossette} {et~al.}(2016){Cossette}, {Charbonneau}, {Smolarkiewic},
  \& {Rast}}]{cossette2016}
{Cossette}, J.~F., {Charbonneau}, P., {Smolarkiewic}, P.~K., \& {Rast}, M.
  2016, Submitted to \apj

\bibitem[{Fan \& Fang(2014)}]{fan2014}
Fan, Y., \& Fang, F. 2014, \apj, 789, 35

\bibitem[{{Field} \& {Blackman}(2002)}]{field2002}
{Field}, G.~B., \& {Blackman}, E.~G. 2002, \apj, 572, 685

\bibitem[{{Ghizaru} {et~al.}(2010){Ghizaru}, {Charbonneau}, \&
  {Smolarkiewicz}}]{ghizaru2010}
{Ghizaru}, M., {Charbonneau}, P., \& {Smolarkiewicz}, P.~K. 2010, \apjl, 715,
  L133

\bibitem[{{Gilman} \& {Rempel}(2005)}]{gilman2005}
{Gilman}, P.~A., \& {Rempel}, M. 2005, \apj, 630, 615

\bibitem[{{Guerrero} {et~al.}(2009){Guerrero}, {Dikpati}, \& {de Gouveia Dal
  Pino}}]{guerrero2009}
{Guerrero}, G., {Dikpati}, M., \& {de Gouveia Dal Pino}, E.~M. 2009, \apj, 701,
  725

\bibitem[{{Hubbard} \& {Brandenburg}(2012)}]{hubbard2012}
{Hubbard}, A., \& {Brandenburg}, A. 2012, \apj, 748, 51

\bibitem[{{K{\"a}pyl{\"a}} {et~al.}(2008){K{\"a}pyl{\"a}}, {Korpi}, \&
  {Brandenburg}}]{kapyla2008}
{K{\"a}pyl{\"a}}, P.~J., {Korpi}, M.~J., \& {Brandenburg}, A. 2008, \aap, 491,
  353

\bibitem[{{K{\"a}pyl{\"a}} {et~al.}(2009){K{\"a}pyl{\"a}}, {Korpi}, \&
  {Brandenburg}}]{kapyla2009}
---. 2009, \aap, 500, 633

\bibitem[{{K{\"a}pyl{\"a}} {et~al.}(2010){K{\"a}pyl{\"a}}, {Korpi},
  {Brandenburg}, {Mitra}, \& {Tavakol}}]{kapyla2010}
{K{\"a}pyl{\"a}}, P.~J., {Korpi}, M.~J., {Brandenburg}, A., {Mitra}, D., \&
  {Tavakol}, R. 2010, \asna, 331, 73

\bibitem[{{K{\"a}pyl{\"a}} {et~al.}(2012){K{\"a}pyl{\"a}}, {Mantere}, \&
  {Brandenburg}}]{kapyla2012}
{K{\"a}pyl{\"a}}, P.~J., {Mantere}, M.~J., \& {Brandenburg}, A. 2012, \apjl,
  755, L22

\bibitem[{{K{\"a}pyl{\"a}} {et~al.}(2013){K{\"a}pyl{\"a}}, {Mantere}, {Cole},
  {Warnecke}, \& {Brandenburg}}]{kapyla2013}
{K{\"a}pyl{\"a}}, P.~J., {Mantere}, M.~J., {Cole}, E., {Warnecke}, J., \&
  {Brandenburg}, A. 2013, \apj, 778, 41

\bibitem[{{Karak} \& {Choudhuri}(2011)}]{karak2011}
{Karak}, B.~B., \& {Choudhuri}, A.~R. 2011, \mnras, 410, 1503

\bibitem[{{Karak} {et~al.}(2014){Karak}, {Rheinhardt}, {Brandenburg},
  {K{\"a}pyl{\"a}}, \& {K{\"a}pyl{\"a}}}]{karak2014}
{Karak}, B.~B., {Rheinhardt}, M., {Brandenburg}, A., {K{\"a}pyl{\"a}}, P.~J.,
  \& {K{\"a}pyl{\"a}}, M.~J. 2014, \apj, 795, 16

\bibitem[{{Kitchatinov} \& {Olemskoy}(2012)}]{kitchatinov2012}
{Kitchatinov}, L.~L., \& {Olemskoy}, S.~V. 2012, \solphys, 276, 3

\bibitem[{{Krause} \& {R\"adler}(1980)}]{krause1980}
{Krause}, F., \& {R\"adler}, K.-H. 1980, {Mean-field magnetohydrodynamics and
  dynamo theory} (Oxford, Pergamon Press, Ltd.)

\bibitem[{{Lawson} {et~al.}(2015){Lawson}, {Strugarek}, \&
  {Charbonneau}}]{lawson2015}
{Lawson}, N., {Strugarek}, A., \& {Charbonneau}, P. 2015, \apj, 813, 95

\bibitem[{{Masada} {et~al.}(2013){Masada}, {Yamada}, \&
  {Kageyama}}]{masada2013}
{Masada}, Y., {Yamada}, K., \& {Kageyama}, A. 2013, \apj, 778, 11

\bibitem[{{Miesch}(2007)}]{miesch2007}
{Miesch}, M.~S. 2007, \apjl, 658, L131

\bibitem[{{Moffatt}(1978)}]{moffatt1978}
{Moffatt}, H.~K. 1978, {Magnetic field generation in electrically conducting
  fluids} (Cambridge University Press, 1978.~353 p., Cambridge)

\bibitem[{{Mu{\~n}oz-Jaramillo} {et~al.}(2011){Mu{\~n}oz-Jaramillo}, {Nandy},
  \& {Martens}}]{munoz-Jaramillo2011}
{Mu{\~n}oz-Jaramillo}, A., {Nandy}, D., \& {Martens}, P.~C.~H. 2011, \apjl,
  727, L23

\bibitem[{{Nelson} {et~al.}(2013){Nelson}, {Brown}, {Brun}, {Miesch}, \&
  {Toomre}}]{nelson2013}
{Nelson}, N.~J., {Brown}, B.~P., {Brun}, A.~S., {Miesch}, M.~S., \& {Toomre},
  J. 2013, \apj, 762, 73

\bibitem[{{Nelson} {et~al.}(2014){Nelson}, {Brown}, {Sacha Brun}, {Miesch}, \&
  {Toomre}}]{nelson2014}
{Nelson}, N.~J., {Brown}, B.~P., {Sacha Brun}, A., {Miesch}, M.~S., \&
  {Toomre}, J. 2014, \solphys, 289, 441

\bibitem[{{Ossendrijver}(2003)}]{ossendrijver2003}
{Ossendrijver}, M. 2003, \aapr, 11, 287

\bibitem[{{Passos} \& {Charbonneau}(2014)}]{passos2014}
{Passos}, D., \& {Charbonneau}, P. 2014, \aap, 568, A113

\bibitem[{{Petrovay}(2010)}]{petrovay2010}
{Petrovay}, K. 2010, \lrsp, 7, 6

\bibitem[{{Pouquet} {et~al.}(1976){Pouquet}, {Frisch}, \&
  {Leorat}}]{pouquet1976}
{Pouquet}, A., {Frisch}, U., \& {Leorat}, J. 1976, \jfm, 77, 321

\bibitem[{{Press} {et~al.}(1992){Press}, {Teukolsky}, {Vetterling}, \&
  {Flannery}}]{press1992}
{Press}, W.~H., {Teukolsky}, S.~A., {Vetterling}, W.~T., \& {Flannery}, B.~P.
  1992, {Numerical recipes in C. The art of scientific computing} (Cambridge:
  University Press, c1992, 2nd ed., Cambridge)

\bibitem[{{Racine} {et~al.}(2011){Racine}, {Charbonneau}, {Ghizaru}, {Bouchat},
  \& {Smolarkiewicz}}]{racine2011}
{Racine}, {\'E}., {Charbonneau}, P., {Ghizaru}, M., {Bouchat}, A., \&
  {Smolarkiewicz}, P.~K. 2011, \apj, 735, 46

\bibitem[{{R\"adler}(1980)}]{radler1980}
{R\"adler}, K.-H. 1980, \asna, 301, 101

\bibitem[{{R\"adler}(2000)}]{radler2000}
{R\"adler}, K.-H. 2000, in Lecture Notes in Physics, Berlin Springer Verlag,
  Vol. 556, From the Sun to the Great Attractor, ed. D.~{Page} \& J.~G.
  {Hirsch} (Springer-Verlag, Berlin Heidelberg), 101--172

\bibitem[{{R{\"a}dler} \& {Stepanov}(2006)}]{radler2006}
{R{\"a}dler}, K.-H., \& {Stepanov}, R. 2006, \pre, 73, 056311

\bibitem[{{R{\"u}diger} {et~al.}(1994){R{\"u}diger}, {Kitchatinov},
  {K{\"u}ker}, \& {Schultz}}]{rudiger1994}
{R{\"u}diger}, G., {Kitchatinov}, L.~L., {K{\"u}ker}, M., \& {Schultz}, M.
  1994, \gafd, 78, 247

\bibitem[{{Schrijver} \& {Siscoe}(2009)}]{schrijver2009}
{Schrijver}, C.~J., \& {Siscoe}, G.~L. 2009, {Heliophysics: Plasma Physics of
  the Local Cosmos} (Cambridge University Press, Cambridge)

\bibitem[{{Schrinner} {et~al.}(2007){Schrinner}, {R{\"a}dler}, {Schmitt},
  {Rheinhardt}, \& {Christensen}}]{schrinner2007}
{Schrinner}, M., {R{\"a}dler}, K.-H., {Schmitt}, D., {Rheinhardt}, M., \&
  {Christensen}, U.~R. 2007, \gafd, 101, 81

\bibitem[{{Simard} {et~al.}(2013){Simard}, {Charbonneau}, \&
  {Bouchat}}]{simard2013}
{Simard}, C., {Charbonneau}, P., \& {Bouchat}, A. 2013, \apj, 768, 16

\bibitem[{{Smolarkiewicz} \& {Charbonneau}(2013)}]{smolarkiewicz2013}
{Smolarkiewicz}, P.~K., \& {Charbonneau}, P. 2013, \jcp, 236, 608

\bibitem[{{Tobias}(1996)}]{tobias1996}
{Tobias}, S.~M. 1996, \apj, 467, 870

\bibitem[{{Vainshtein} \& {Cattaneo}(1992)}]{vainshtein1992}
{Vainshtein}, S.~I., \& {Cattaneo}, F. 1992, \apj, 393, 165

\bibitem[{{Warnecke} {et~al.}(2016){Warnecke}, {Rheinhardt}, {K{\"a}pyl{\"a}},
  {K{\"a}pyl{\"a}}, \& {Brandenburg}}]{warnecke2016}
{Warnecke}, J., {Rheinhardt}, M., {K{\"a}pyl{\"a}}, P.~J., {K{\"a}pyl{\"a}},
  M.~J., \& {Brandenburg}, A. 2016, ArXiv e-prints

\bibitem[{{Weiss}(2010)}]{weiss2010}
{Weiss}, N. 2010, \ageo, 51, 9

\end{thebibliography}

\end{document}